\documentclass[10pt]{article} 
\ifdefined\AnonymousSubmission
  \usepackage{tmlr}
\else
  \usepackage[preprint]{tmlr}
\fi

\usepackage{amsmath,amsfonts,bm}

\def\eqref#1{equation~\ref{#1}}

\def\1{\bm{1}}

\DeclareMathAlphabet{\mathsfit}{\encodingdefault}{\sfdefault}{m}{sl}
\SetMathAlphabet{\mathsfit}{bold}{\encodingdefault}{\sfdefault}{bx}{n}

\usepackage{amsthm}
\newtheorem{proposition}{Proposition}
\usepackage{graphicx}
\usepackage{booktabs}
\usepackage{array}
\newcommand{\appendixtableformat}{%
  \setlength{\abovecaptionskip}{0pt}%
  \setlength{\belowcaptionskip}{6pt}%
  \setlength{\tabcolsep}{4pt}%
  \renewcommand{\arraystretch}{1.08}%
}
\usepackage{placeins}
\usepackage{xcolor}
\makeatletter
\newcommand{\appendixtableband}[1]{%
  \noalign{%
    \begingroup
    \color{black!6}%
    \vbox to 0pt{%
      \hrule width\linewidth height
        \dimexpr#1\ht\@arstrutbox+#1\dp\@arstrutbox\relax
      \vss
    }%
    \endgroup
    \nointerlineskip
  }%
}
\makeatother
\graphicspath{{figures/}}

\usepackage{hyperref}
\usepackage{url}

\title{
On Task Scope and Information Retention in Source Coding 
}

\author{%
  \name Alireza Furutanpey\thanks{Corresponding author} \email a.furutanpey@neverblink.eu \\
  \addr NeverBlink, Warszawa, Poland
  \AND
  \name Kerstin Bunte \email k.bunte@rug.nl \\
  \addr Rijksuniversiteit Groningen, Groningen, Netherlands
}

\begin{document}
\raggedbottom

\maketitle

\begin{abstract}
We argue that dividing codec design into \emph{Coding for Machines (CfM)} and \emph{Coding for Humans (CfH)} is a misleading distinction for deciding what information a codec may discard. Receiver identity does not determine admissible information loss. The required rate depends on task scope, including the predictions to support, their losses and tolerated risks, the encoder observation, and the permitted decoding procedures. Notably, a machine task may have a higher minimum rate than a restricted human decision. Rate savings on selected machine tasks apply only to the stated requirements, not to an intrinsic ordering by receiver type. We extend source and feature coding to finite task families, derive when restricting the encoder observation preserves the minimum rate, and show that equality between source and split-feature coding rates can no longer hold as the task scope expands.%
\ifdefined\AnonymousSubmission
\footnote{Our use of generative AI was limited to language editing, manuscript consistency checks, and coding assistance. The authors take responsibility for the final text, claims, results, and code implementation.}%
\fi
\end{abstract}

\section{Introduction} \label{sec:introduction}
The line of work in Coding for Machines (CfM) studies compression when transmitted data are used primarily for downstream machine analysis \citep{duan2020vcm,choi2022scalable,feng2022omnipotent,shao2022task,gunduz2023beyond,harell2025rd}. A common motivation is the growing volume of data that must be transmitted over bandwidth-limited links in mobile applications and remote sensing. The corresponding system model assumes a resource-constrained client that continuously acquires high-dimensional data and transmits a compact representation to a better-provisioned remote server shared by multiple clients. The server performs prediction tasks and may return or relay their results. A cloud provider offering a feature codec to application developers may not know their tasks at codec design time. Such a service must specify the task scope it supports, since savings on designer-selected tasks do not establish suitability for its clients. Compression must offset codec overhead and, in time-critical applications, reduce end-to-end request latency. At the information-theoretic level, CfM is a source-coding problem in which fidelity is specified by the information that must be preserved to satisfy downstream prediction-risk requirements.

While CfM may provide a sound formal model for theoretical analysis, we argue that the implied division of codec design into \emph{Coding for Machines} and \emph{Coding for Humans} is misleading, particularly as a basis for deciding which source information a codec should preserve to reduce rate. 

Receiver identity alone provides no indication of which source distinctions may be discarded. Rather, the required rate depends primarily on the task scope, i.e., what must be predicted, to what tolerance, and subject to what observation and decoding constraints.
Therefore, two machine tasks may require different source information, while a restricted human decision may require less information than either faithful reconstruction or a machine task with broader predictive requirements.

Task-specific rate savings establish an advantage only for the stated requirements and do not imply a general ordering of minimum rates by receiver type. A machine task may instead have the \textit{higher minimum rate}. Consider that prediction models can exploit source distinctions that human observers cannot reliably perceive \citep{ilyas2019adversarial,geirhos2019imagenet,furutanpey2025fool}. If those distinctions reduce prediction risk, preserving that lower risk forces the coded message to retain information that a human-limited task may discard, i.e., a model with better prediction performance through such information is likely to require more bits to preserve that performance. Moreover, reconstruction serves different purposes for general human viewing and for expert inspection of a prediction. The former may require broad perceptual fidelity, whereas the latter requires only that the evidence relevant to the decision remain interpretable. If the coded message preserves that evidence, the same message can support a human-interpretable reconstruction without introducing a separate reconstruction channel. Source details irrelevant to the decision need not be preserved solely because a human views the reconstruction.
A simple preliminary experiment with two machine tasks illustrates this dependence on task scope (Figure~\ref{fig:pilot}). 

\begingroup
\setlength{\intextsep}{10pt}
\begin{figure}[htb]
\centering
\includegraphics[width=\linewidth]{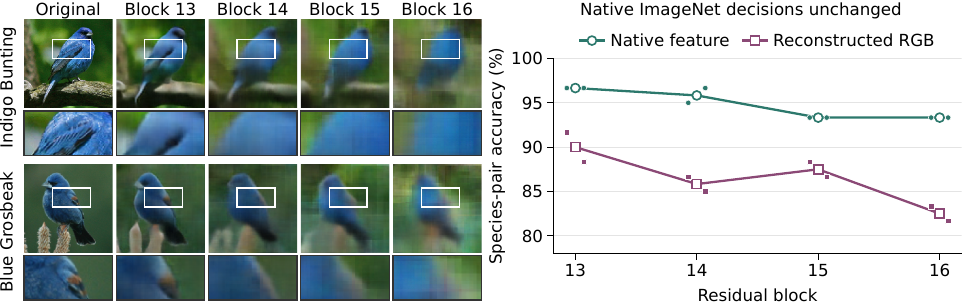}
\caption{
ImageNet ResNet-50 inversions for Indigo Bunting (top) and Blue Grosbeak (bottom). Boxes mark the wing regions enlarged below. The chestnut patch becomes less visible at later blocks. Species-pair accuracy uses 200-way CUB readers restricted to these species, with 30 test photographs each. Filled points show two fits per observation and open points their means (Appendix~\ref{app:pilot}).
}
\label{fig:pilot}
\end{figure}
\endgroup

The frozen remaining layers of the ImageNet-pretrained ResNet-50 reproduce the original ImageNet classification decisions at every shown block. The color cue distinguishing Indigo Bunting from Blue Grosbeak becomes less visible in the later reconstruction. At block 16, species-pair accuracy is $93.3\%$ from native features and $82.5\%$ from reconstructed RGB, averaged over the fitted readers \citep{wah2011cub}. The original ImageNet decisions remain unchanged, while species-pair accuracy differs between prediction from native features and prediction from reconstructed RGB. A reconstruction constraint does not remove this dependence on task scope because it preserves only the distinctions penalized by its distortion criterion and tolerance. Guaranteeing human verification outside the stated task scope requires preserving distinctions relevant to those additional inspections. The listed task-risk constraints do not, in general, guarantee their preservation when those inspections are unspecified.
This is precisely the setting that motivates CfM. Yet, the CfM/CfH distinction risks imposing a misleading design prior, directing attention toward receiver identity rather than the task scope that ultimately determines what information may be discarded.
\emph{Why, then, should codec design be divided into Coding for Machines and Coding for Humans when admissible information loss is determined by the supported task scope rather than receiver identity}?

Through theoretical analysis and empirical studies, we show how extrapolating coding equivalence beyond the original task can mislead codec design. We extend a recent rate--distortion formulation of coding for machines \citep{harell2025rd} to finite task families with uncertain reference outputs. We characterize when a reduced encoder observation preserves the coding optimum and bound the penalty when it does not. Coding from a feature can match source coding for the original task yet require more bits when another task is added, despite unchanged optimal uncompressed predictions. Learned finite codes demonstrate this penalty in completely transmitted messages. Image-compression studies further show how preservation losses and decoding procedures affect the predictions obtainable from a message. All code and reported results are openly available in \url{https://github.com/rezafuru/Task-Scope-and-Information-Retention}.

\section{Task scope and compression requirements} \label{sec:rel_work}
\emph{Task scope} specifies the required predictions, their losses and tolerated risks, the encoder observation, and the permitted joint predictions. Reconstruction requirements and decoder choices also restrict message reuse.
\subsection{Task scope and rate--distortion theory} \label{sec:background}
Let $X$ be a source variable and $\widehat X$ a decoded output. We evaluate their discrepancy with a distortion function $d(x,\hat x)$ and restrict $\widehat X$ to a reconstruction alphabet $\widehat{\mathcal X}$, which may contain images, features, or predictions. For independent and identically distributed samples on finite alphabets, additive distortion, and unrestricted block coding, the minimum asymptotic rate is
\begin{equation}
R(D)
=
\min_{p(\hat x\mid x)}
\left\{I(X;\widehat X)\,\middle|\,\mathbb E[d(X,\widehat X)]\le D\right\},
\label{eq:classical-rate-distortion}
\end{equation}
where $D$ is the tolerated expected distortion, and mutual information is measured in bits. Sufficiently long block codes can approach this rate. Two compression problems with the same source and receiver can have different minimum rates when they impose different reconstruction alphabets or distortion requirements. Accordingly, a rate comparison must specify what the decoded message must preserve.

To represent several predictions from the same message, we model $X$ jointly with finite-alphabet reference outputs $T_1,\ldots,T_m$. The index $q\in Q=\{1,\ldots,m\}$ identifies a task, and $T_q$ denotes its reference output. For a task asking whether an image contains a vehicle, $T_q$ is the binary reference label. A reference output may remain uncertain even when $X$ is known, as with noisy labels. Lowercase $x$ and $t_q$ denote their realizations. 

The encoder may have access either to the full source $X$ or to a restricted observation $Y=g(X)$, where $g$ is a fixed observation or feature map applied before coding and is not adapted to the particular task $q$. 
We denote the encoder's 
\emph{available
} observation by $O$. 
For a block of $n$ independent samples, the encoder sends one message $M_n=e_n(O^n)$ at rate $R$ bits per sample. Decoder $q$ computes $\widehat T_q^n=\psi_{q,n}(M_n)$ from that same message. We assume no source-correlated side information at the decoders. 
Restricting $O$ through $g$ can merge source states before coding. If $g$ merges states with different conditional target distributions to the same observation, the smallest attainable prediction risk can increase. A fixed viewing interface for a human operator can impose the same restriction when it renders two measurements indistinguishable despite their different evidence for the task. A predictor observing $X$ can then predict with lower risk, and preserving that risk can require a higher rate.

For a single sample, 
let $\widehat T=(\widehat T_1,\ldots,\widehat T_m)$ denote 
the joint decoded prediction and let $\hat t$ exhibit a particular prediction tuple. 
We fix a finite, nonempty set $\widehat{\mathcal T}$ of permitted tuples. Separate task decoders may permit 
every combination of individual predictions, whereas 
prediction functions applied to one common reconstruction may allow only a subset of combinations. All channels, sums, and optimizations over $\hat t$ are taken over $\widehat{\mathcal T}$ 
which need not equal the set of possible reference-output tuples.

We evaluate $\hat t_q$ against $t_q$ using a task distortion $d_{T_q}(t_q,\hat t_q)\in[0,1]$. 
Its expectation is the prediction risk. 
For each task, we require $\mathbb{E}[d_{T_q}(T_q,\widehat T_q)]\leq D_q$, and 
write $\mathbf D=(D_1,\ldots,D_m)$ for the vector of task tolerances. 
Because the target $T_q$ 
may remain uncertain even when the 
observation is fixed at $O=o$, we average the distortion over its conditional distribution and define 
\begin{equation}
c_{q,O}(o,\hat t)
=
\mathbb{E}\!\left[d_{T_q}(T_q,\hat t_q)\mid O=o\right].
\label{eq:indirect-distortion}
\end{equation}
A prediction channel based on $O$ satisfies
$\mathbb{E}[d_{T_q}(T_q,\widehat T_q)]
= \mathbb{E}[c_{q,O}(O,\widehat T)].$
We apply indirect rate-distortion theory with one expected-loss constraint for each required task \citep{witsenhausen1980indirect,ivry2026answer}. The minimum asymptotic rate is
\begin{equation}
R_O(\mathbf D;Q)
=
\min_{p(\hat t\mid o)}
\left\{
I(O;\widehat T)
\,\middle|\,
\mathbb{E}[d_{T_q}(T_q,\widehat T_q)]\leq D_q
\text{ for every } q\in Q
\right\}.
\label{eq:task-family-rate}
\end{equation}
We set the rate to infinity when no permitted prediction channel satisfies all requirements and omit $Q$ from the notation when the family is fixed.

The Bayes risk $D^{\star}_{q,O}$ is the smallest expected task loss achievable from observation $O$. If $D_q<D^{\star}_{q,O}$, the risk requirement cannot be achieved from $O$ at any rate. We measure excess risk relative to the Bayes risk for a specified reference observation, $X$ or $Y$. A feature $Y$ is sufficient for target $T_q$ when $T_q$ and $X$ are conditionally independent given $Y$. The same conditional target distribution is then available from either observation, so either permits an optimal prediction for any loss \citep{blackwell1953equivalent,vanrooyen2014lecam}.

For a fixed loss, preserving an optimal prediction can require fewer distinctions than preserving the posterior. Assuming binary zero-one loss, posterior probabilities $0.6$ and $0.9$ for class~1 both give the Bayes prediction $1$. Predicting $0$ instead increases conditional error by $0.2$ and $0.8$, respectively. An encoder that distinguishes these states can place more coding errors on the state with posterior $0.6$, where each error has lower cost. An observation that merges the states removes that choice. Multiple tasks can further restrict the error allocation. If another task assigns high error cost to states that are inexpensive for the first task, satisfying both requirements can require a higher rate even when both tasks share the same Bayes prediction.

\subsection{Related Work}
\paragraph{Receiver labels and mixed fidelity requirements.}
Researchers in coding for machines, supervised compression, and task-oriented semantic communication specify fidelity through selected inference objectives \citep{duan2020vcm,matsubara2022supervised,matsubara2023sc2,shao2022task,gunduz2023beyond}. \citet{zhang2026machines} motivate their framework by describing machine vision as requiring less information than human vision while evaluating selected detection, segmentation, and reconstruction objectives. Task-specific compression can also preserve human decisions at reduced rates \citep{reddy2021pragmatic}. 
JPEG AI specifies a shared representation for visualization, image processing, and computer vision \citep{jpegai-requirements}.
\citet{stavrou2023fidelity} analyze joint semantic and observation reconstruction subject to separate fidelity constraints. 
\paragraph{Task families and preserved decisions.}
\citet{dubois2021lossy} formulate preservation through excess prediction risk over task families defined by shared invariances. \citet{ivry2026answer} derives a finite-family rate expression with one tuple of query-specific predictions and separate excess-risk constraints. For a fixed loss, \citet{sevetlidis2026bayes} characterizes representations that preserve Bayes risk by recovering an optimal prediction. When a statistic preserves the target posterior, \citet{armstrong2026source} proves information-bottleneck and log-loss equivalence through conditional averaging. \citet{wang2026predictive} obtain equal source and posterior coding rates when distortion depends on the source through its target posterior. Optimal predictions can agree across tasks even when coding errors have different conditional costs.
\paragraph{Coding interfaces and conditional error costs.}
\citet{choi2022scalable} compare source and feature reconstructions given a prescribed task distortion and discuss the possible inadequacy of a trained representation for later tasks. \citet{harell2025rd} add direct-feature coding and evaluate distortion against a fixed model output $T=f(X)=h(Y)$. 
They give output conditions for rate equalities. 
With multiple targets, each may remain uncertain given $X$ and need not be determined by $Y$. \citet{martinian2008distortion} study source coding with observed distortion costs, and \citet{enttsel2026model} compare indirect coding with compressing a model's prediction. Finite Shannon optimization and conditions on the predictions used by an optimum are established in rational inattention \citep{caplin2019rational,armenter2024geometry}.
\paragraph{Reusable learned representations.}
Researchers learn features for reuse across vision tasks and exploit dependencies between task representations during coding \citep{feng2022omnipotent,guo2025tasks,huang2026adaptive}. Other methods use common and task-specific messages \citep{andrade2026graywyner}, auxiliary reconstruction for secondary tasks \citep{andrade2024compatible}, task-derived importance weights \citep{esfahanizadeh2026unitac}, or transformed features from different tasks and architectures \citep{gao2025dtufc}. With FrankenSplit and FOOL, researchers compress shallow features and evaluate their reuse across downstream models \citep{furutanpey2024frankensplit,furutanpey2025fool}. Reusing a codec across tasks, obtaining an additional prediction from an existing message, and fitting a new decoder after coding impose different requirements. 
Rate comparisons must state which of these requirements are included.

\section{Comparing source and feature coding}
\label{sec:coding-comparison}

An encoder observing $X$ can use source distinctions that are absent from $Y=g(X)$. We compare three coding arrangements with identical task-risk tolerances and permitted prediction tuples.

In source coding, we encode $X$ and predict from a source reconstruction $\widehat X\in\widehat{\mathcal X}$. In split-feature coding, we encode $Y$ and predict from a feature reconstruction $\widehat Y$. In direct-feature coding, we encode $X$ and predict from a feature reconstruction $\widetilde Y$. We denote their minimum rates by $R_X$, $R_Y$, and $R_{XY}$, respectively. Both feature reconstructions take values in $\widehat{\mathcal Y}$.

\subsection{Jointly attainable predictions}

Let $F(\hat x)=(f_1(\hat x),\ldots,f_m(\hat x))$ and $H(\hat y)=(h_1(\hat y),\ldots,h_m(\hat y))$ be fixed prediction functions 
of the source and feature reconstructions, respectively. Suppose their finite reconstruction alphabets satisfy
\begin{equation}
F(\widehat{\mathcal X})=H(\widehat{\mathcal Y}).
\label{eq:joint-output-range}
\end{equation}
We use this common range as the permitted prediction tuples in \eqref{eq:task-family-rate}. Every permitted tuple is obtainable from both reconstruction alphabets.
Matching the output values available for each task separately does not ensure that every combination of those outputs 
is attainable from a single reconstruction.

At task tolerances $\mathbf D$, we obtain each rate by minimizing the corresponding mutual information,
\begin{equation}
\begin{aligned}
I(X;\widehat X),&\qquad X\longrightarrow\widehat X\xrightarrow{F} \widehat T,
&&R_X(\mathbf D;Q),\\
I(Y;\widehat Y),&\qquad Y\longrightarrow\widehat Y\xrightarrow{H} \widehat T,
&&R_Y(\mathbf D;Q),\\
I(X;\widetilde Y),&\qquad X\longrightarrow\widetilde Y\xrightarrow{H} \widehat T,
&&R_{XY}(\mathbf D;Q),
\end{aligned}
\label{eq:interface-rates}
\end{equation}
over reconstruction channels satisfying $\mathbb E[d_{T_q}(T_q,\widehat T_q)]\le D_q$ for every task.

Replacing a reconstruction with its prediction tuple preserves all task risks and cannot increase mutual information. Conversely, by \eqref{eq:joint-output-range}, we can choose one representative reconstruction for each permitted tuple. From any prediction channel, we can reconstruct that representative with the same mutual information and risks. We can therefore compute the three rates in \eqref{eq:interface-rates} using the prediction optimization in \eqref{eq:task-family-rate}, with $O=X$ for source and direct-feature coding and $O=Y$ for split-feature coding (Appendix~\ref{app:joint-outputs}).

\subsection{Encoder observation and task risk}
\label{sec:observation-risk}

For an input $x$ and prediction tuple $\hat t$, write $c_q(x,\hat t)=c_{q,X}(x,\hat t)$ for the conditional task distortion defined in \eqref{eq:indirect-distortion}. Averaging over inputs with the same feature gives
\begin{equation}
\bar c_q(y,\hat t)=c_{q,Y}(y,\hat t)=\mathbb E[c_q(X,\hat t)\mid Y=y].
\label{eq:conditional-feature-loss}
\end{equation}
An encoder observing $X$ can compute $Y$, so $R_{XY}(\mathbf D)=R_X(\mathbf D)\le R_Y(\mathbf D)$. For Harell's target $T=f(X)=h(Y)$, every conditional cost depends on $X$ only through $Y$. 
Averaging a source prediction channel over inputs with the same feature preserves task distortion and cannot increase rate. 
Thus $R_{XY}=R_Y$, and $R_X$ is also equal when \eqref{eq:joint-output-range} holds \citep{harell2025rd}. 
For additional targets, the conditional costs of 
different predictions can vary among inputs with the same feature. 
Whether those differences affect the minimum rate depends on the tolerated risks.

\subsection{Exact equality and rate penalty}
\label{sec:exact-observation}

For finite nonnegative multipliers $\boldsymbol\lambda$, write $c_{\boldsymbol\lambda,O}=\sum_q\lambda_qc_{q,O}$ and define
\begin{equation}
J_O(\boldsymbol\lambda)=\min_{p(\hat t\mid o)}
\left\{I(O;\widehat T)+\mathbb E c_{\boldsymbol\lambda,O}(O,\widehat T)\right\}.
\label{eq:weighted-rate-objective}
\end{equation}
Each $\lambda_q$ is the marginal cost of increasing task $q$'s risk.
For a distribution $r$ over permitted prediction tuples, the standard rate-distortion variational formula gives the conditional distribution
\begin{equation}
\pi_O^r(\hat t\mid o)
=\frac{r(\hat t)2^{-c_{\boldsymbol\lambda,O}(o,\hat t)}}
{\sum_{\hat u}r(\hat u)2^{-c_{\boldsymbol\lambda,O}(o,\hat u)}}.
\label{eq:gibbs-prediction}
\end{equation}
At an optimum, $r$ is the marginal distribution of the decoded prediction. The variational reduction and its support conditions are 
given in \citep{caplin2019rational,armenter2024geometry}. Comparing the two observations at the same $r$ gives
\begin{equation}
G_{\boldsymbol\lambda}(r)
=\mathbb E_X D_{\mathrm{KL},2}\!\left(
\pi_Y^r(\cdot\mid Y)\,\Vert\,\pi_X^r(\cdot\mid X)\right),
\label{eq:conditional-gibbs-gap}
\end{equation}
where $D_{\mathrm{KL},2}$ uses base-two logarithms. 
This quantity measures the divergence between 
the prediction channels caused by averaging the conditional costs within each feature value.

\begin{proposition}[Prescribed observation]
\label{prop:exact-observation}
For the finite model, assume \eqref{eq:joint-output-range}. Then $J_X(\boldsymbol\lambda)=J_Y(\boldsymbol\lambda)$ if and only if some globally optimal source marginal $r$ satisfies
\begin{equation}
c_{\boldsymbol\lambda,X}(x,\hat t)-c_{\boldsymbol\lambda,X}(x,\hat u)
=c_{\boldsymbol\lambda,X}(x',\hat t)-c_{\boldsymbol\lambda,X}(x',\hat u)
\label{eq:cost-contrast-criterion}
\end{equation}
whenever $g(x)=g(x')$, $p_X(x)p_X(x')>0$, and $r(\hat t)r(\hat u)>0$. 
For any globally optimal marginals $r_X,r_Y$ for 
the respective objectives,
\begin{equation}
G_{\boldsymbol\lambda}(r_Y)
\le J_Y(\boldsymbol\lambda)-J_X(\boldsymbol\lambda)
\le G_{\boldsymbol\lambda}(r_X).
\label{eq:objective-gap-bounds}
\end{equation}
\end{proposition}

The equality condition involves only cost differences between predictions with positive probability in an optimal channel. Adding the same source-dependent cost to every used prediction leaves their relative costs unchanged. Posteriors can therefore vary among inputs with the same feature value while the minimum objective values still remain equal. Identical Bayes predictions alone do not ensure equality, because they do not specify the relative costs of coding-errors. 
The predictions with positive probability in an optimal channel can change with the tolerance. Requiring a global optimum excludes arbitrary singleton supports, which satisfy \eqref{eq:cost-contrast-criterion} vacuously (Appendix~\ref{app:exact-observation}).

The rate penalty at the same task-risk tolerances follows by evaluating the two objectives at their respective dual-optimal multipliers. If $\mathbf D$ is feasible from $Y$ and finite multipliers $\boldsymbol\lambda_X,\boldsymbol\lambda_Y$ attain the respective constrained dual optima, then
\begin{equation}
\Delta J(\boldsymbol\lambda_X)
\le R_Y(\mathbf D)-R_X(\mathbf D)
\le\Delta J(\boldsymbol\lambda_Y),
\qquad \Delta J=J_Y-J_X.
\label{eq:matched-risk-gap}
\end{equation}
Strict feasibility of the risk constraints suffices for 
attainment by finite multipliers. 
Equal multipliers need not include 
equal task risks. Combining \eqref{eq:objective-gap-bounds} and \eqref{eq:matched-risk-gap} gives quantitative bounds 
on the rate penalty at the required componentwise tolerances. 
For Harell's fixed target, the conditional costs satisfy \eqref{eq:cost-contrast-criterion} on the entire prediction range. 
The conditional costs for an expanded task family can violate this criterion even when the original task and its risk tolerance 
remain unchanged.

The encoder's observation and the training objective must also be specified separately, since the latter does not by itself determine the information retained by the encoder. 
When a source-observing encoder is trained using distortion against $T=f(X)$, it can still retain 
distinctions between inputs with the same $T$. 
We therefore cannot infer the Markov chain $X^n\longrightarrow T^n\longrightarrow M_n$ from that training objective alone. 
When the distortion depends on $X$ only through $T$, we can construct a target-dependent channel at the unrestricted optimum by averaging the encoder channel over $X$ given $T$.
A fitted encoder need not implement this target-dependent 
channel, even if it achieves the same training risk.

\section{Finite families at nonzero risk}
\label{sec:task-families}

\subsection{Noisy threshold tasks}
\label{sec:threshold-family}

Let $S$ be uniform on $\{0,\ldots,5\}$, let $X=(S,N)$ include an independent finite-valued sensor variation $N$, and let $Y=\lfloor S/2\rfloor$. We require predictions of two noisy threshold labels. The underlying decisions are
\begin{equation}
C=\mathbf 1\{S\geq 2\},\qquad B=\mathbf 1\{S\geq 3\}.
\label{eq:threshold-decisions}
\end{equation}
We denote task indices as $C$ and $B$, with tolerances ordered as $\mathbf D=(D_C,D_B)$. The targets are $T_C=C\oplus E_C$ and $T_B=B\oplus E_B$, where $\oplus$ adds bits modulo two, flipping the threshold bit when $E_q=1$. 
The noise variables $E_C$, $E_B$ are independent Bernoulli$\epsilon$ variables with parameter $\epsilon\in(0,1/2)$, independent of $S$. 
The sensor variation $N$ is independent of $(S,E_C,E_B)$. Both tasks use zero--one loss, $d_{T_q}(t_q,\hat t_q)=\mathbf 1\{t_q\ne \hat t_q\}$ for $q\in\{C,B\}$.

The source threshold functions and the feature outputs $h_C(y)=\mathbf 1\{y\geq 1\}$ and $h_B(y)=\mathbf 1\{y\geq 2\}$ permit exactly the same joint predictions $(\widehat T_C,\widehat T_B)\in\{(0,0),(1,0),(1,1)\}$.
The prediction $(0,1)$ is excluded because the higher threshold cannot be positive when the lower threshold is negative. Independent label noise still permits the reference-output tuple $(T_C,T_B)=(0,1)$.

We can recover $C$ from $Y$. Recovering $B$ without error requires distinguishing $S=2$ from $S=3$, which share $Y=1$. These equally likely values occur with total probability $1/3$, giving minimum underlying-bit error $1/6$ from $Y$. If a decoded prediction has underlying-bit error $e_q$, its target risk is $\epsilon+(1-2\epsilon)e_q$. The split-observation Bayes risk for the additional task is therefore $D_{B,Y}^*=\epsilon+(1-2\epsilon)/6$. Both full-observation Bayes risks are $\epsilon$.

At the original task's Bayes risk, every code must recover $C$, requiring $H_C=h_2(2/3)$ bits per sample, where $h_2$ is binary entropy. At or above the reduced-observation risk floor, transmitting $C$ also satisfies the additional task 
requirement. Below that floor, source-observing encoders additionally encode $B$ within the event 
$C=1$, which has 
probability $2/3$. The independent sensor variation $N$ can be discarded. No code based on the reduced observation has additional-task risk below this floor. Proposition~\ref{prop:thresholds} in Appendix~\ref{app:thresholds} gives the exact rates and conditional binary-coding construction.

\begin{figure}[ht]
\centering
\includegraphics[width=\linewidth]{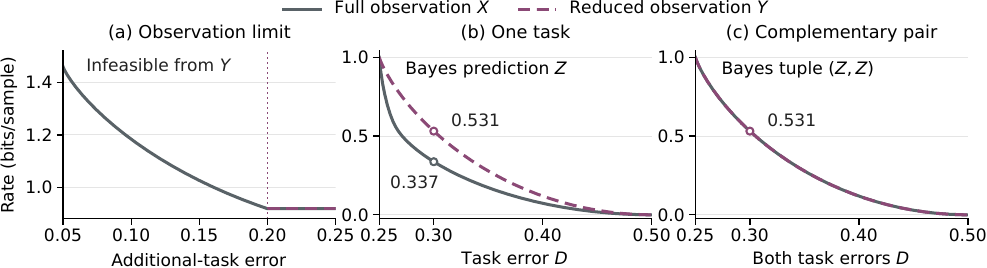}
\caption{Exact asymptotic rates for two noisy task families. (a) Label noise and original-task error are $0.05$. Minimum additional-task error from $Y$ is $0.20$. At or above this floor, the rates coincide. (b) At noise probabilities $(0.05,0.45)$, both observations have Bayes prediction $Z$ and minimum error $0.25$, with different rates at intermediate tolerances. (c) Reversing these probabilities for a complementary task makes the rates equal at $D_1=D_2=D$, with Bayes tuple $(Z,Z)$. Circles mark $D=0.30$. Single-task curves specialize \citet{martinian2008distortion}'s formula. The pair rate is \eqref{eq:confidence-complementary-rate}.
}
\label{fig:exact-limits}
\end{figure}

Keeping encoder access fixed at $X$, we can compare requirements specified by the reference observations $X$ and $V=Y$. We keep the original-task tolerance at $\epsilon$ and add the same nonnegative excess-risk allowance to each reference's additional-task Bayes risk. 
With the coarser reference, the minimum rate is $H_C$ bits per sample. 
With $X$ as reference, the minimum rate is higher while this allowance is less than $(1-2\epsilon)/6$ (Figure~\ref{fig:exact-limits}a). This increment is smaller than $H(X\mid V)=1+H(N)$ because the independent sensor variation remains irrelevant to both tasks. Appendix~\ref{app:thresholds} gives the rate difference.

Replacing $B$ by $B'=\mathbf 1\{S\geq4\}$ retains two tasks and makes both underlying decisions deterministic functions of $Y$. Encoding their joint Bayes predictions requires $\log_2 3=1.58496$ bits per sample, compared with $1.45915$ bits per sample for $(C,B)$. Using $Y$, we can attain the full-observation Bayes risks of this higher-rate family. We cannot attain those of the lower-rate family from $Y$. Task count and the minimum rate from $X$ therefore do not determine whether a particular reduced observation preserves the required predictions.

\subsection{Retaining the original task while adding a requirement}
\label{sec:nonbinary-family}

A feature can preserve the original task's entire rate-distortion function while increasing the minimum rate required by an expanded task family. Let $Z\in\{0,1,2\}$ have probabilities $(0.35,0.35,0.30)$, let $K\in\{-1,1\}$ be an independent fair sign, and set $X=(Z,K)$ and $Y=Z$. 
The original task is the noiseless target $T_0=Z$. 
An additional noisy ternary target $T_1$ has the unique Bayes-optimal prediction $Z$ from either observation and Bayes risk $0.45$. Both tasks use zero--one loss.

For $Z=0$ or $1$, predicting the other of these two classes has conditional excess risk 
$0.325$, irrespective of $K$. Predicting class $2$ instead increases risk by $0.325+0.27625K$. 
The encoder observing $K$ can therefore allocate some prediction errors 
to states where its cost is $0.04875$, rather than $0.60125$. 
At $Z=2$, either wrong prediction has conditional excess cost $0.325$. Appendix~\ref{app:nonbinary} gives the complete target law. 
The minimum Bayes margin is $0.04875$, so the Bayes prediction is unique.

For the original task alone, all three coding arrangements achieve 
the same minimum rate-distortion function. 
Keep its tolerance at $D_0=0.35$ and add $D_1=0.499$. The resulting family rates are
\begin{equation}
R_X(0.35,0.499)=0.63540,
\qquad R_Y(0.35,0.499)=0.81876
\label{eq:nonbinary-family-rates}
\end{equation}
bits per sample. The optimal source and feature channels have original-task risks 
$0.29456$ and $0.15077$, respectively. Both satisfy the unchanged original requirement. The gap persists when all nine prediction tuples are permitted. The additional task alone yields a lower bound attained by setting $\widehat T_0=\widehat T_1$ in these feasible channels.

The gap is $0.18336$ bits per sample, lying within the interval $[0.16642,0.19660]$ from \eqref{eq:matched-risk-gap}. 
With exact original-task recovery, $D_0=0$, both encoders instead require $H(Z)$ bits per sample and can attain the additional task Bayes risk $0.45$ by transmitting $Z$. 
Thus the penalty depends on the allowed risks 
as well as the added task. 
A feature selected using the original fixed-target equality can restrict 
the encoder's ability to 
allocate 
errors to states with lower conditional excess 
costs, even without changing either task's best uncompressed prediction.

\subsection{Task-dependent error costs}
\label{sec:confidence-family}

Complementary tasks can eliminate 
an encoder's rate advantage from observing confidence. 
Let $Z$ and $K$ be independent fair bits, set $X=(Z,K,N)$ and $Y=Z$, and let $N$ be finite-valued sensor variation 
independent of all task variables. The two zero--one tasks have targets $T_q=Z\oplus E_q$. 
Conditional noise probabilities in states $K=0,1$ are $(\epsilon_0,\epsilon_1)$ for task 1 and reversed for task 2, independently of $Z$, with $0\le\epsilon_0<\epsilon_1<1/2$. Both Bayes predictions are $Z$, with risk $p=(\epsilon_0+\epsilon_1)/2$. 
The readouts $f_q(z,k,n)=z$ and $h_q(y)=y$ therefore yield the same 
decoded bit $U$ for both tasks.

Write $w_k=1-2\epsilon_k$ and $e_k=\Pr(U\ne Z\mid K=k)$. The tasks' excess risks are $(w_0e_0+w_1e_1)/2$ and $(w_1e_0+w_0e_1)/2$. 
For task 1 alone, observing $K$ permits a larger error probability 
in the less costly state $K=1$, as in distortion-side-information coding \citep{martinian2008distortion}. 
An encoder observing only $Y$ cannot condition its error probability on $K$, so 
$e_0=e_1$.

At a common tolerance $D$, averaging the two risk constraints gives $(e_0+e_1)/2\le(D-p)/(1-2p)$. 
Averaging the coding channel over $N$ and symmetrizing it with its bit-complemented counterpart, preserves the risks and cannot increase the rate, giving rate $1-[h_2(e_0)+h_2(e_1)]/2$. 
Concavity of binary entropy then implies that the 
constant error allocation is optimal. Encoders observing $X$ or $Y$ can use this constant error allocation, so for $p\le D\le1/2$,
\begin{equation}
R_X^{(1,2)}(D,D)=R_Y^{(1,2)}(D,D)
=1-h_2\!\left(\frac{D-p}{1-2p}\right).
\label{eq:confidence-complementary-rate}
\end{equation}
The sum of the two conditional excess costs is 
independent of $K$, satisfying \eqref{eq:cost-contrast-criterion}. Proposition~\ref{prop:confidence} in Appendix~\ref{app:confidence} extends this criterion to arbitrary finite confidence states and task families, including unrestricted joint predictions and unequal tolerances.

For $(\epsilon_0,\epsilon_1)=(0.05,0.45)$ and $D=0.30$, task 1 alone requires $0.33680$ bits per sample from $X$ and $0.53100$ bits per sample from $Y$. The 
source-optimal channel for task 1 alone gives task 2 risk $0.44486$. 
Requiring both risks to be at most $0.30$ raises the minimum source rate to $0.53100$ (Figure~\ref{fig:exact-limits}b,c). The Bayes tuple remains $(Z,Z)$, yet 
the rate advantage of observing confidence disappears once both tasks are imposed. 
Counting optimal predictions or comparing uncompressed accuracy therefore does not determine the rate required when prediction errors are allowed. 

\section{Conditional error costs in finite coding}
\label{sec:learned-allocation}

We fit three codebooks for the three-class family, each with 8,192 ternary words of length 16. 
Both decoders return the codeword indexed by the transmitted message. 
Full and reduced selectors use either the known conditional costs or estimates of these costs fitted from independent noisy labels. We evaluate the codes on fresh source blocks using the known target law, with tolerances $D_0=0.35$ and $D_1=0.499$. We count framing and padding bits when computing complete rates (Appendix~\ref{app:nonbinary-codes}).

\begin{figure}[tbp]
\centering
\includegraphics[width=\linewidth]{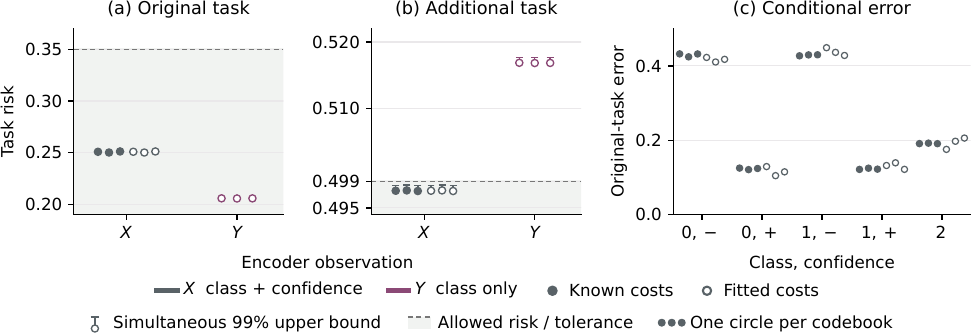}
\caption{Finite codes at $0.81251$ bits per symbol, below the $0.81876$ minimum for satisfying both requirements from $Y$. (a,b) Codes whose encoders observe class and confidence satisfy both requirements. (c) Their errors concentrate in negative-confidence states, where they incur less additional-task loss.}
\label{fig:nonbinary-codes}
\end{figure}

For all three full-observation codes, the simultaneous one-sided 99\% Hoeffding upper risk bounds are below their respective task tolerances (Figure~\ref{fig:nonbinary-codes}a,b). Their complete rate of $0.81251$ bits per source symbol is below the $0.81876$-bit analytical minimum rate for 
reduced-observation coding under 
both requirements, irrespective of how the selector is fitted 
(\eqref{eq:nonbinary-family-rates}). The strict rate advantage of observing the full source in 
Section~\ref{sec:nonbinary-family} therefore persists for finite codebooks, 
including framing and padding.

The full encoder allocates more 
original-task errors to the 
negative-confidence states, where predicting class 2 incurs less additional-task loss (Figure~\ref{fig:nonbinary-codes}c). 
Selecting codewords using 
fitted conditional costs likewise allocates more 
original-task errors in the negative-confidence states and satisfies both risk requirements. On the same codebooks, reduced-observation selectors cannot distinguish the confidence states, and their additional-task risks exceed the tolerance. 
Preserving the Bayes prediction therefore does not preserve the available allocation of coding errors across confidence states. 

Two controls preserve the 
class distribution and Bayes predictions. With only the original deterministic task, full and reduced selection give identical packets. 
Full and reduced selection likewise produce 
identical packets when 
the additional task's target law is averaged over confidence. 
Observing confidence changes these codes only when the 
conditional error costs relevant to the coding objective vary with confidence. 
In the binary family, tighter complementary requirements eliminate the benefit of 
concentrating 
errors in either state. The single-task saving disappears when corruption makes the observed confidence independent of the underlying state, i.e. 
at $\tau=1/2$ (Appendix~\ref{app:block-coding}). These effects follow Proposition~\ref{prop:confidence}.

\section{Requirements for reusable image codes}
\label{sec:experiments}

Image-code rates are measured in bits per pixel (bpp) using complete messages, including the main latent, hyperlatent, and header. All reported predictions are computed from entropy-decoded messages. Shared model parameters are stored separately (Appendix~\ref{app:empirical}). 
Classification requirements allow at most three percentage points of coarse-accuracy loss and five points of fine-accuracy loss relative to 
uncompressed references.

\subsection{Coarse supervision and fine prediction}
\label{sec:imagenet}

ImageNet's ENTITY-30 grouping has 30 superclasses and 240 fine classes \citep{santurkar2021breeds}. Codecs fitted with either objective encode the same layer-2 tensor from an ImageNet-pretrained ResNet-50 at $224\times224$ resolution and reconstruct it for the frozen suffix. 
Starting from a 
shared feature-reconstruction initialization, 
coarse-only fitting minimizes KL between original and decoded coarse probabilities plus rate. 
Added-fine fitting additionally minimizes 
normalized fine-label cross entropy, holding the architecture, observation, coarse-loss and rate terms, and fitting exposure fixed (Appendix~\ref{app:imagenet-methods}).

We use validation data to select readout routes and minimum mean complete rate among 
paired codec fits. 
We compare 
minimum rates under the 
coarse-only and joint mean accuracy-drop requirements within the same candidate pool, and retain a 
selection satisfying both requirements for each fitting objective.

\begin{figure}[t]
\centering
\includegraphics[width=\linewidth]{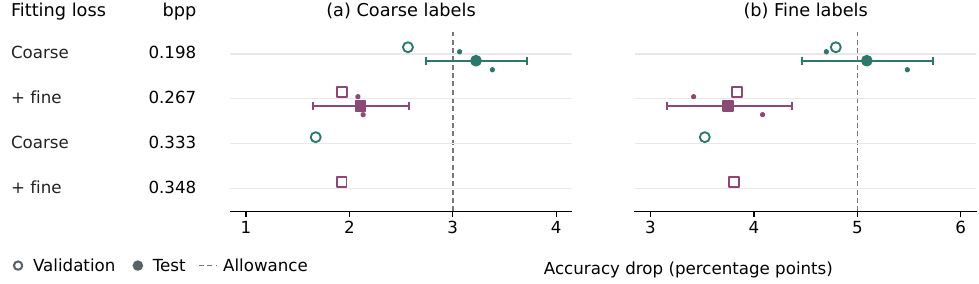}
\caption{ImageNet accuracy drops from uncompressed references. Rows list fitting losses and complete validation rates. Open and filled markers show validation and selected test means. Small dots show individual fits, and bars show conditional 95\% paired-image intervals. Dashed lines mark the allowances.}
\label{fig:imagenet-scope}
\end{figure}

Coarse-only and joint validation selections coincide. On 6,000 test images, the selected coarse-only codec's coarse- and fine-accuracy drops exceed their respective allowed losses 
(Fig.~\ref{fig:imagenet-scope}). 
With added-fine coding, the simultaneous 95\% upper limits on both accuracy drops are below their allowances, at a complete rate of $0.26717$ bpp, compared with $0.19772$ bpp for the coarse-only objective. 
The original objective and encoder observation alone do not determine whether an added task's risk requirement will be 
satisfied. Both codecs observe the same early tensor, so this experiment isolates the effect of
the fitted preservation objective rather than the observation restriction studied in Section~\ref{sec:observation-risk}.

\subsection{Preservation and adapted decoding}
\label{sec:cifar}

CIFAR-100 has 20 coarse and 100 fine classes \citep{krizhevsky2009learning}. 
A ResNet-18 \citep{he2016residual} teacher is trained on coarse labels. 
The teacher's early $128\times16\times16$ feature tensor serves as 
the encoder input. 
A tensor with 
the same dimensions is reconstructed from each 
message. 
The preservation loss is either 
squared error on the early features or squared error on the 20 coarse logits produced by 
the frozen teacher suffix. Fine labels are not used in either preservation loss. 

Coarse predictions are computed with the frozen suffix. 
The inherited fine classifier is kept 
fixed after training on uncompressed early tensors. 
Adapted classifiers are refitted using 
decoded tensors or reconstructed logits, respectively. 
The coarse and fine accuracy-drop allowances are the same as for ImageNet, with fine predictions obtained from 
refitted tensor readouts. For each of three teachers and each preservation loss, the lowest-rate paired candidate satisfying both mean requirements is fixed before test assessment (Appendix~\ref{app:cifar-methods}).

\begin{figure}[t]
\centering
\includegraphics[width=\linewidth]{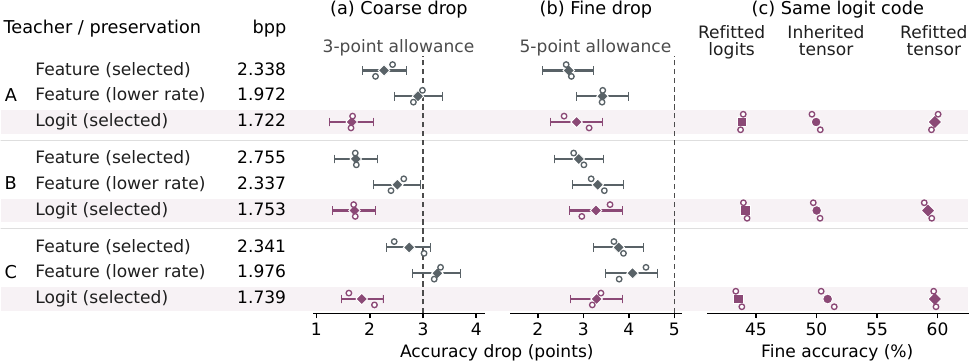}
\caption{Test results for validation-selected CIFAR codes and nearby lower-rate feature-preserving candidates. (a,b) Accuracy drops from uncompressed references, with dashed allowance lines. (c) Fine readouts applied to each unchanged logit-preserving message. In (a,b), diamonds show means over paired codec fits, hollow circles show individual runs, and bars show conditional 95\% paired-image intervals.}
\label{fig:cifar-local}
\end{figure}

Across teachers, the selected logit-preserving codes achieve 
smaller coarse-accuracy drops at lower complete rates than the selected feature-preserving codes (Fig.~\ref{fig:cifar-local}a). Refitted fine accuracies are similar (Fig.~\ref{fig:cifar-local}b). 
The 95\% intervals for the selected logit-minus-feature differences fall 
within the one-percentage-point equivalence margin, conditional on the fitted models (Appendix~\ref{app:cifar-methods}). 
For the nearby feature-preserving candidates at lower-rates, 
rates remain higher than those of 
the logit-preserving selections and coarse-accuracy drops approach or exceed the allowance. 
Teacher C's selected feature-preserving group includes one codec fit that failed, 
and its interval for coarse-drop 
crosses the allowance threshold.

Refitting the fine classifier on decoded tensors improves fine 
accuracy without changing the logit-preserving message (Fig.~\ref{fig:cifar-local}c). 
The classifier refitted on reconstructed logits is less accurate 
than either tensor readout. 
Therefore, the inherited classifier's errors do not establish that the message lacks information useful for distinguishing fine labels. 
The encoder input remains the early tensor, and a loss on coarse logits alone does not require the message to retain information 
only about 
those logits (Section~\ref{sec:observation-risk}). 
Fine distinctions can therefore remain recoverable 
without fine-label supervision during codec fitting, provided the decoder is adapted to the coded representation.

Measured rates characterize 
tested codecs and fitted readouts. Appendix~\ref{app:residual-information} compares prediction-only transmission and the rate needed to limit residual information.

\subsection{Task losses and decoder choices}
\label{sec:taskonomy}

\begin{figure}[t]
\centering
\includegraphics[width=\linewidth]{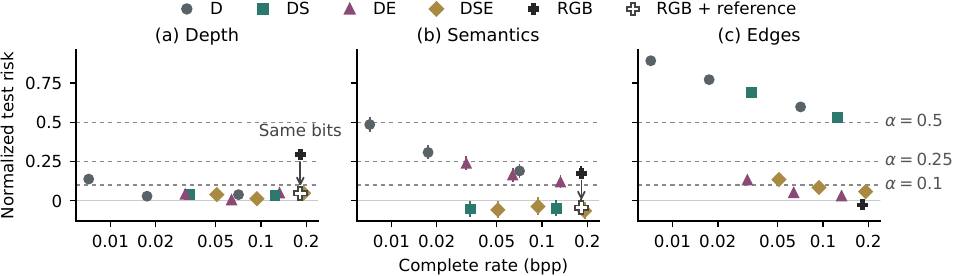}
\caption{Normalized Taskonomy test risks for twelve codecs. Filled symbols show risks with matched task decoders. Arrows connect risks to those obtained from the same RGB-preserving message by reconstructing RGB and applying frozen reference predictors (hollow symbols). The latent edge decoder is unchanged. Dashed lines mark tolerated risks, and bars show conditional 95\% image-bootstrap intervals.}
\label{fig:taskonomy-requirements}
\end{figure}

Taskonomy contains aligned indoor RGB images, depth maps, and semantic pseudo-labels \citep{zamir2018taskonomy}. 
The same RGB inputs are encoded with codecs fitted using combinations of depth ($D$), semantic ($S$), and RGB-derived edge ($E$) losses, or with RGB reconstruction loss. 
The matched task decoders in Fig.~\ref{fig:taskonomy-requirements} are fitted after freezing each codec, 
with a common architecture, matched training exposure, and the same 
validation checkpoint rule.

We evaluate log-depth Smooth-L1 loss, weighted semantic cross entropy, and edge mean squared error. Stringent, moderate, and loose requirements permit at most 
10\%, 25\%, and 50\%, respectively, of the loss increase from each validation reference to its constant predictor. 
The thresholds remain fixed when evaluated on 2,000 test images from five unseen buildings. 
We plot risks using 
the same normalization in Figure~\ref{fig:taskonomy-requirements} and assess feasibility from 
point estimates. Appendix~\ref{app:taskonomy-methods} reports 
the corresponding absolute tolerances.

Near $0.03$ bpp, the depth-semantic and depth--edge codes have similar depth risks (Fig.~\ref{fig:taskonomy-requirements}a). 
Semantic loss is lower for 
depth-semantic coding, and edge loss is lower for 
depth-edge coding (Fig.~\ref{fig:taskonomy-requirements}b,c). 
The displayed depth-edge candidate satisfies all three moderate requirements without semantic supervision in 
codec fitting. 
For decoders selected using the available-decoder rule, the mean risks over three fits also satisfy these requirements, although only one 
individual fit satisfies 
all three (Appendix~\ref{app:taskonomy-choices}). 
Hence, the task losses used for codec fitting do not specify the complete set of predictions obtainable from the resulting 
message.

Reconstructing RGB and applying the frozen reference predictors reduces 
depth and semantic risks without changing the transmitted message, as indicated by the vertical arrows in Fig.~\ref{fig:taskonomy-requirements}a,b. 
Depth risk falls from above the moderate threshold to below the stringent threshold when the frozen reference predictor is applied to reconstructed RGB. All three stringent requirements are satisfied when the latent edge decoder is retained. 

All three fits of the $D,30$ depth-only setting satisfy the loose semantic cross-entropy requirement. 
Their mean object 
intersection-over-union (mIoU) is below $0.01$. 
In the validation examples, depth-only \mbox{semantic} maps omit labeled objects or assign them incorrect categories (Fig.~\ref{fig:taskonomy-examples}). 
Therefore, a probability-loss requirement alone does not establish categorical agreement with the reference labels. 
We set the additional moderate mIoU floor at $0.75$ times the validation reference 
mIoU. 
The uncompressed reference's test mIoU falls 
below this floor. Its mIoU exceeds the floor in a retrospective diagnostic restricted to classes present in the validation set, thereby excluding the newly appearing test class (Appendix~\ref{app:semantic-support}). 
Categorical requirements must specify both the metric and the set of classes to be evaluated.

\section{Conclusion}
Receiver identity alone does not determine which source information a codec may discard.
We extended source and feature coding to finite task families, characterized when restricting encoder observation leaves the minimum rate unchanged, and bounded the resulting rate penalty.
Our theoretical constructions and learned finite codes demonstrate that restricting observation can preserve the minimum rate for one task while increasing it for an expanded task family.

The image studies show that information useful for additional predictions can remain available after fitting with a narrower loss. With adapted decoding, we can satisfy additional risk requirements without changing the transmitted message or its rate.
Compression savings must therefore be interpreted relative to the specified predictions, losses, tolerances, encoder observations, and decoding constraints.
Preserving each task's optimal uncompressed prediction need not preserve the distinctions required to allocate coding errors efficiently under the tasks' risk constraints.

\bibliography{main}
\bibliographystyle{tmlr}

\appendix
\section{Coding results and proofs}
\label{app:theory}

The coding model has finite alphabets, independent and identically distributed samples, componentwise losses, and the permitted prediction tuples of Section~\ref{sec:background}.

\subsection{Jointly attainable predictions and equality of coding rates}
\label{app:joint-outputs}

For a source reconstruction channel, define $\widehat T=F(\widehat X)$. Every task risk is unchanged, and data processing gives $I(X;\widehat T)\le I(X;\widehat X)$. Given \eqref{eq:joint-output-range}, choose a representative $s(\hat t)\in\widehat{\mathcal X}$ with $F(s(\hat t))=\hat t$ for every permitted prediction tuple $\hat t$. Given a prediction channel, set $\widehat X=s(\widehat T)$. Because the representative and its prediction tuple determine one another, $I(X;\widehat X)=I(X;\widehat T)$, with the same task risks. Applying the same construction with representatives in $\widehat{\mathcal Y}$ proves the reduction for the split and direct rates.

For the deterministic target $T=f(X)=h(g(X))$ in \citet{harell2025rd}, the conditional expected task distortion depends on $X$ only through $Y=g(X)$. Averaging a source-dependent reconstruction channel over $X$ given $Y$ preserves that distortion and does not increase mutual information. Hence the direct and split rates are equal. Source equality additionally uses an attainable-output condition or a replacement output with no greater distortion. For several tasks, one replacement must have no greater distortion for any task. Separate task replacements may form a tuple that no reconstruction can produce.

\subsection{Prescribed observation}
\label{app:exact-observation}

Fix $\boldsymbol\lambda$. All sums range over the permitted prediction tuples in \eqref{eq:joint-output-range}. For $O=X,Y$, put
\begin{equation}
Z_O^r(o)=\sum_{\hat t}r(\hat t)2^{-c_{\boldsymbol\lambda,O}(o,\hat t)},
\qquad
\Phi_O(r)=-\mathbb E\log_2 Z_O^r(O).
\label{eq:observation-partition}
\end{equation}
Finite costs make $Z_O^r$ positive throughout the compact probability simplex. For any prediction channel $p$ with marginal $p_{\widehat T}$,
\begin{align}
\mathbb E D_{\mathrm{KL},2}(p(\cdot\mid O)\Vert r)
&=I(O;\widehat T)+D_{\mathrm{KL},2}(p_{\widehat T}\Vert r),\notag\\
\mathbb E D_{\mathrm{KL},2}(p(\cdot\mid O)\Vert r)
+\mathbb E c_{\boldsymbol\lambda,O}(O,\widehat T)
&=\Phi_O(r)+\mathbb E D_{\mathrm{KL},2}(p(\cdot\mid O)\Vert\pi_O^r(\cdot\mid O)).
\label{eq:observation-variational}
\end{align}
Minimizing jointly over $p,r$ gives $J_O=\min_r\Phi_O(r)$. At a global minimum, $\pi_O^r$ has marginal $r$. These are the finite Shannon variational identities underlying the support conditions of \citet{caplin2019rational,armenter2024geometry}.

On the support of $r$, the logarithm of the ratio $\pi_Y^r(\hat t\mid y)/\pi_X^r(\hat t\mid x)$ equals
\[
c_{\boldsymbol\lambda,X}(x,\hat t)-c_{\boldsymbol\lambda,Y}(y,\hat t)
+\log_2 Z_X^r(x)-\log_2 Z_Y^r(y).
\]
Average with probabilities $p_X(x)\pi_Y^r(\hat t\mid g(x))$. Conditional expectation given $Y$ cancels the cost difference, proving
\begin{equation}
\Phi_Y(r)-\Phi_X(r)=G_{\boldsymbol\lambda}(r).
\label{eq:observation-kl-identity}
\end{equation}
Evaluating this identity at $r_Y$ and $r_X$ proves \eqref{eq:objective-gap-bounds}.

If \eqref{eq:cost-contrast-criterion} holds at an $X$-optimal $r$, then on its support $c_{\boldsymbol\lambda,X}(x,\hat t)=u(x)+v(g(x),\hat t)$. The factor $2^{-u(x)}$ cancels in \eqref{eq:gibbs-prediction}, so $\pi_X^r=\pi_Y^r$ and $J_X=J_Y$. Conversely, equality gives $J_X\le\Phi_X(r_Y)\le\Phi_Y(r_Y)=J_Y$. Hence $r_Y$ is also $X$-optimal and its two Gibbs channels coincide. For supported tuples, the ratio of the Gibbs probabilities satisfies
\[
\frac{\pi_X^{r_Y}(\hat t\mid x)}{\pi_X^{r_Y}(\hat u\mid x)}
=\frac{r_Y(\hat t)}{r_Y(\hat u)}
2^{-[c_{\boldsymbol\lambda,X}(x,\hat t)-c_{\boldsymbol\lambda,X}(x,\hat u)]},
\]
which proves \eqref{eq:cost-contrast-criterion}. With finite costs, the Gibbs probabilities are positive on the support of $r_Y$, so these ratios are defined. Global optimality additionally requires
\begin{equation}
\mathbb E_X\frac{2^{-c_{\boldsymbol\lambda,X}(X,\hat t)}}{Z_X^r(X)}\le1
\quad\text{for every permitted }\hat t,
\qquad\text{with equality if }r(\hat t)>0.
\label{eq:observation-unused-predictions}
\end{equation}
$G_{\boldsymbol\lambda}(r_Y)$ can equal zero when $J_Y>J_X$, because $J_Y-J_X=G_{\boldsymbol\lambda}(r_Y)+\Phi_X(r_Y)-J_X$.

For \eqref{eq:matched-risk-gap}, use $R_O(\mathbf D)=J_O(\boldsymbol\lambda_O)-\boldsymbol\lambda_O\cdot\mathbf D$. Evaluating the $Y$ dual at $\boldsymbol\lambda_X$ gives the lower bound, and evaluating the $X$ dual at $\boldsymbol\lambda_Y$ gives the upper bound. At any $\mathbf D$ feasible from $Y$, the rates are equal exactly when some constrained $X$-optimal prediction channel factors through $Y$. If the rates are equal, a $Y$-optimal channel composed with $g$ is also $X$-optimal. Conversely, an $X$-optimal channel that depends on $X$ only through $Y$ is feasible from $Y$ at the same rate. If finite source multipliers $\boldsymbol\lambda_X$ attain the dual optimum, an equivalent certificate is an $X$-optimal Gibbs marginal $r$ at these multipliers satisfying \eqref{eq:cost-contrast-criterion}, whose channel has risks $D_q^r\le D_q$ and $\lambda_{X,q}(D_q-D_q^r)=0$ for every $q$. At feasible boundaries, compactness gives $R_O(\mathbf D+\epsilon\mathbf1)\to R_O(\mathbf D)$ as $\epsilon\downarrow0$. Applying the finite-multiplier bounds at these strictly feasible relaxed tolerances provides boundary bounds by limits. If $\mathbf D$ is infeasible from $Y$, then $R_Y(\mathbf D)=+\infty$.

\subsection{Three-class conditional error costs}
\label{app:nonbinary}

Let $P(Z)=(0.35,0.35,0.30)$, let $K$ be an independent fair sign, and set $X=(Z,K)$, $Y=Z$. The original target is $T_0=Z$. Write $a=0.55$, $b=0.225$, $h=a-b=0.325$, $\rho=0.85h$, and $s=K\rho/3$. The additional target has conditional law
\begin{equation}
\begin{aligned}
P(T_1=\cdot\mid Z=0,K)&=(a+s,b+s,b-2s),\\
P(T_1=\cdot\mid Z=1,K)&=(b+s,a+s,b-2s),\\
P(T_1=\cdot\mid Z=2,K)&=(b,b,a).
\end{aligned}
\label{eq:nonbinary-reference-law}
\end{equation}
For zero--one loss, both observations have unique Bayes tuple $(Z,Z)$ and risks $(0,0.45)$. Every conditional class probability is positive, and the minimum Bayes margin is $h-\rho=0.04875$.

For the additional task alone, put $\beta=\lambda h$. Symmetry and convexity allow a marginal with $r(0)=r(1)=(1-r(2))/2$. At $r(2)=0$, the derivative of $\Phi_X$ with respect to $r(2)$ is $(1-S_X)/\ln2$, where
\begin{equation}
S_X(\beta)=0.7\frac{2^{-0.15\beta}+2^{-1.85\beta}}{1+2^{-\beta}}+0.3\,2^\beta.
\label{eq:nonbinary-entry}
\end{equation}
For $\beta>0$, prediction $2$ is unused exactly when $S_X\le1$. For $t=2^\beta$, multiplying $S_X-1$ by $t+1$ gives $f(t)=0.3t^2-0.7t-1+0.7(t^{0.85}+t^{-0.85})$. On $t\ge1$, $f(1)=0$, $f'(1)=-0.1$, and $f''(t)\ge0.51075$. Since $f(t)\to\infty$, it has exactly one further root. Prediction $2$ has positive probability in the source-optimal marginal for $\beta>\beta_X\simeq0.18152990$. Averaging the costs replaces $t^{0.85}+t^{-0.85}$ by $2$. Prediction $2$ has positive probability in the feature-optimal marginal for $\beta>\beta_Y=\log_2(4/3)\simeq0.41503750$.

For $\beta>0$, the exponential-cost matrices have full column rank, so their marginal objectives are strictly convex. Divide each source row by its correct-prediction entry. Subtracting the exchanged class rows forces the first two coordinates of a null vector to agree. Subtracting the confidence-sign rows then forces its third coordinate to vanish, and the first two vanish as well. The feature matrix has diagonal $1$ and off-diagonal $2^{-\beta}$. For $0<\beta\le\beta_X$, only predictions $0,1$ have positive probability in the source-optimal marginal. Their conditional cost difference is independent of $K$, so $J_X=J_Y$. For $\beta>\beta_X$, the source-optimal marginal assigns positive probability to prediction $2$ and at least one other prediction. Their conditional cost difference depends on $K$, so $J_X<J_Y$. A singleton prediction-$2$ optimum is excluded by its larger constant-prediction risk.

At the finite-code confirmation's tolerance $D_1=0.499$, minimizing the strictly convex marginal objective gives source multiplier $\lambda_X\simeq8.936033247655$, rate $0.635401148729$, and original-label error $0.294561875080$. Writing $p=(0.35,0.35,0.30)$ and $e=(D_1-0.45)/0.325$, Fano's inequality for a ternary variable gives a matching lower bound for
\begin{equation}
R_Y^{(1)}(D_1)=H(p)-h_2(e)-e,\qquad 0\le e\le0.6.
\label{eq:nonbinary-feature-rate}
\end{equation}
To attain it, choose $r_i=(p_i-e/2)/(1-3e/2)$ and the backward channel $P(Z=i\mid\widehat T_1=j)=1-e$ for $i=j$, and $e/2$ otherwise. Mixing these conditional distributions with weights $r_j$ gives marginal $p$ for $Z$ and conditional entropy $H(Z\mid\widehat T_1)=h_2(e)+e$. At $D_1=0.499$, the feature rate is $0.818759706492$ and its original-label error is $e\simeq0.150769230769$.

Set $D_0=0.35$ and permit all nine prediction pairs. Projection onto $\widehat T_1$ gives the standalone additional-task rate as a family lower bound. Setting $\widehat T_0=\widehat T_1$ attains it from either observation while satisfying $D_0$. Thus the family-rate gap at $(D_0,D_1)=(0.35,0.499)$ is approximately $0.183358557763$ bits per sample.

The interval of equal rates for the additional task alone ends near $0.00199467$ bits per sample. For the common-prediction construction on this interval, the original-task error is $0.30+0.70/(1+2^\beta)\ge0.62800929$, above $D_0=0.35$. Even with separate predictions, the original-task Fano bound is $H(p)-h_2(0.35)-0.35\simeq0.29722284$. Finally, if $D_0=0$, every feasible message determines $Z$ and requires at least $H(Z)$ bits per sample. Sending $Z$ attains additional risk $0.45$, so both family rates equal $H(Z)$ for every feasible $D_1\ge0.45$.

\subsection{Information beyond a preservation target}
\label{app:residual-information}

The chain-rule argument of \citet[Proposition 3.1]{achille2018invariance} bounds information beyond a deterministic preservation target. With lossy target reconstruction, the bound depends on complete coded length and the target's rate--distortion function.

Let $(X_i,U_i)$ be iid and finite, and let $T_i=f(X_i)$. A complete binary prefix-coded message $M$ is generated from $X^n$ using randomness independent of the source and targets. A decoder using only $M$ reconstructs $T^n$ at mean additive distortion at most $D$. Write $r=\mathbb E|M|/n$ and let $R_T(D)$ be the ordinary target rate--distortion function assuming the same distortion and reconstruction alphabet. Then
\begin{equation}
\frac1n I(U^n;M\mid T^n)\le r-R_T(D).
\label{eq:residual-information}
\end{equation}
By the prefix-code converse, the chain rule, the target rate--distortion converse, and conditional data processing,
\begin{align}
nr&\ge H(M)\ge I(X^n;M)\notag\\
&=I(T^n;M)+I(X^n;M\mid T^n)\notag\\
&\ge nR_T(D)+I(U^n;M\mid T^n).
\end{align}
The last step uses $U^n\longrightarrow X^n\longrightarrow M$ and deterministic $T^n$. Source-independent shared model parameters can be conditioned on throughout. A rate close to $R_T(D)$ limits additional information beyond $T^n$. The image-code rates are not compared with an estimated $R_T(D)$.

If only the known coarse and fine prediction tuple is required, computing both predictions at the encoder permits transmission in 11 fixed-length bits per image before framing.

\subsection{Noisy threshold tasks}
\label{app:finite-families}
\label{app:thresholds}

\begin{proposition}[Overlapping thresholds]
\label{prop:thresholds}
Fix $D_C=\epsilon$, write $e=(D_B-\epsilon)/(1-2\epsilon)$, and let $H_C=h_2(2/3)$, where $h_2(u)=-u\log_2u-(1-u)\log_2(1-u)$ is binary entropy, with $0\log_2 0=0$. For $D_B\geq\epsilon$,
\begin{equation}
R_X(\epsilon,D_B)=R_{XY}(\epsilon,D_B)=
\begin{cases}
H_C+\dfrac{2}{3}\bigl[h_2(1/4)-h_2(3e/2)\bigr],&e<1/6,\\[3pt]
H_C,&e\geq1/6.
\end{cases}
\label{eq:threshold-rate}
\end{equation}
The split-feature rate $R_Y(\epsilon,D_B)$ is infinite below $D_{B,Y}^*$ and equals $H_C$ at or above it. All three rates are infinite if either task requires risk below $\epsilon$.
\end{proposition}

For Proposition~\ref{prop:thresholds}, independent label noise gives target risk $\epsilon+(1-2\epsilon)e_q$, where $e_q$ is the error in predicting underlying threshold bit $q\in\{C,B\}$. For the target risk to converge to $\epsilon$, the average error in $C$ must converge to zero. The pair $(C,B)$ has probabilities $(1/3,1/6,1/2)$ on $(00,10,11)$. On the event $C=0$, $B=0$. Conditional on $C=1$, $B$ is Bernoulli with parameter $3/4$.

For a length-$n$ encoded message $M_n$ and average error $e_C$ in the decoded coarse predictions, binary Fano's inequality gives
\begin{equation}
I(C^n;M_n)\geq nH_C-nh_2(e_C).
\label{eq:threshold-coarse-converse}
\end{equation}
If the additional threshold has average error $e_B<1/6$, its error conditional on $C=1$ is at most $3e_B/2$. The conditional entropy bound and concavity of binary entropy give
\begin{equation}
I(B^n;M_n\mid C^n)
\geq\frac{2n}{3}\bigl[h_2(1/4)-h_2(3e_B/2)\bigr].
\label{eq:threshold-fine-converse}
\end{equation}
For this bound, expand $H(B^n\mid M_n,C^n)$ over positions, retain only the conditioning on $C_i$ and decoded prediction $\widehat T_{B,i}$, and apply binary Fano's inequality when $C_i=1$. The total error on those positions is at most the overall error. Averaging over positions gives Eq.~\ref{eq:threshold-fine-converse}. Since both $C^n$ and $B^n$ are functions of $X^n$, the sum of these two mutual informations is at most $I(X^n;M_n)$. Letting $e_C\to0$ proves the converse in Eq.~\ref{eq:threshold-rate}. When $e_B\geq1/6$, the coarse bound alone gives $H_C$.

For achievability, encode $C^n$ losslessly at asymptotic rate $H_C$ and apply a binary Hamming rate--distortion code to the subsequence with $C_i=1$. Its rate per original sample is the second term in Eq.~\ref{eq:threshold-rate}. For $e_B\geq1/6$, send only $C$ and predict $\widehat T_B=C$. The three joint predictions are obtained from source representatives $\widehat S=0,2,3$, with any fixed permitted value of $\widehat N$, or feature representatives $\widetilde Y=0,1,2$. Independent $N$ affects neither task's distortion and can be discarded.

The feature value $Y=1$ corresponds to equally likely $S=2,3$, so every prediction of $B$ from $Y$ has underlying-bit error at least $(1/3)(1/2)=1/6$. At or above the resulting target-risk floor, transmitting $C$ attains rate $H_C$, which the coarse task also requires. This proves the split-feature formula.

We can use the same family to compare reference prediction risks while keeping encoder access fixed at $X$. Let $V=Y$ be a coarser observation used only to specify those risks, and let $\eta\geq0$ be the tolerated excess risk for the additional task. With tolerances equal to the reference risks from $V$ plus allowance $(0,\eta)$, the minimum rate is $H_C$ bits per sample. Using the reference risks from $X$ with the same allowance requires an additional
\begin{equation}
\Delta R(\eta)=\frac{2}{3}\left[h_2(1/4)-h_2\!\left(\frac{3\eta}{2(1-2\epsilon)}\right)\right]
\label{eq:threshold-reference-gap}
\end{equation}
for $0\leq\eta<(1-2\epsilon)/6$, and zero thereafter. At $\epsilon=0.05$ and zero allowance, the difference is $0.54085$ bits per sample (Figure~\ref{fig:exact-limits}a). Preserving the lower errors for these two noisy decisions requires fewer additional bits than $H(X\mid V)=1+H(N)$, which includes the independent sensor variation.

With encoder observation fixed at $X$, the reference risk vectors are $(\epsilon,\epsilon)$ from $X$ and $(\epsilon,\epsilon+(1-2\epsilon)/6)$ from $V=Y$. Adding $(0,\eta)$ to each and substituting into Eq.~\ref{eq:threshold-rate} gives Eq.~\ref{eq:threshold-reference-gap}.

For arbitrary tolerances on both decisions, we obtain the complete region by solving the finite-alphabet rate--distortion problem with two constraints. Order the prediction columns as $(00,10,11)$. Index the source rows by the underlying pairs $(C,B)=(00,10,11)$ with probabilities $(1/3,1/6,1/2)$, and the feature rows by $Y=0,1,2$ with probabilities $(1/3,1/3,1/3)$. The corresponding underlying-bit distortion matrices are
\begin{equation}
d_C=
\begin{pmatrix}
0&1&1\\
1&0&0\\
1&0&0
\end{pmatrix},\qquad
d_{B,X}=
\begin{pmatrix}
0&0&1\\
0&0&1\\
1&1&0
\end{pmatrix},\qquad
d_{B,Y}=
\begin{pmatrix}
0&0&1\\
1/2&1/2&1/2\\
1&1&0
\end{pmatrix}.
\label{eq:threshold-distortion-matrices}
\end{equation}
The coarse-task matrix is the same for both observations. For the noisy targets, transform every matrix entry $d$ into $\epsilon+(1-2\epsilon)d$ to obtain the conditional expected target loss \citep{witsenhausen1980indirect,kipnis2015binary}.

\subsection{Conditional error profiles}
\label{app:confidence}

Let $Z$ be a fair bit independent of finite $K\in\mathcal K$, with $\Pr(K=k)>0$ for every $k$. Set $X=(Z,K,N)$ and $Y=Z$, with finite sensor variation $N$ independent of all task variables. Each task has target $T_q=Z\oplus E_q$, zero--one loss, and
\[
\Pr(E_q=1\mid Z,K=k)=\epsilon_q(k)<1/2.
\]
An encoder observing $X$ has access to $K$ but not $E_q$. Every task has the unique Bayes prediction $Z$ from either observation. Its Bayes risk and conditional excess cost for predicting $1-Z$ are
\begin{equation}
p_q=\mathbb E\epsilon_q(K),\qquad
w_q(k)=1-2\epsilon_q(k)>0,\qquad
\bar w_q=\mathbb Ew_q(K)=1-2p_q.
\label{eq:confidence-costs}
\end{equation}
Observing $K$ permits more coding errors where $w_q(k)$ is smaller, as in distortion-side-information coding \citep{martinian2008distortion}.

For a tolerated risk vector $\mathbf D=(D_1,\ldots,D_m)$ with $D_q\geq p_q$, define
\begin{equation}
e^*=\min\left\{\frac12,\min_{q\in Q}\frac{D_q-p_q}{\bar w_q}\right\},
\qquad
\mathbf v_q=\left(\frac{w_q(k)}{\bar w_q}\right)_{k\in\mathcal K}.
\label{eq:confidence-profiles}
\end{equation}
Here $e^*$ is the largest common crossover probability in $[0,1/2]$ that satisfies all risk constraints when the encoder observes only $Z$. The vector $\mathbf v_q$ records task $q$'s normalized error costs across confidence states and has expectation one over $K$. Let $J\subseteq Q$ contain exactly the tasks attaining the inner minimum, and let $\mathbf1$ denote the vector of ones indexed by $\mathcal K$.

\begin{proposition}[Conditional error profiles]
\label{prop:confidence}
Allow every joint prediction in $\{0,1\}^m$. The minimum rates for observations $X$ and $Y$ are
\begin{align}
R_X(\mathbf D)&=1-\max_e\mathbb E\bigl[h_2(e(K))\bigr],
\label{eq:confidence-source-rate}\\
R_Y(\mathbf D)&=1-h_2(e^*).
\label{eq:confidence-feature-rate}
\end{align}
We maximize over functions $e\colon\mathcal K\to[0,1/2]$ satisfying
\[
\mathbb E\bigl[w_q(K)e(K)\bigr]\leq D_q-p_q,
\qquad q\in Q.
\]
Here $e(k)$ is the conditional probability of predicting $1-Z$ at state $k$. Each minimum rate is attained by a channel predicting the same bit for every task, although all joint prediction tuples are permitted. For $0<e^*<1/2$,
\begin{equation}
R_X(\mathbf D)=R_Y(\mathbf D)
\quad\Longleftrightarrow\quad
\mathbf 1\in\operatorname{conv}\{\mathbf v_q:q\in J\}.
\label{eq:confidence-criterion}
\end{equation}
The convex hull contains all weighted averages with nonnegative weights summing to one. If the condition fails, $R_X<R_Y$. Both rates are infinite if any $D_q<p_q$, one if some $D_q=p_q$ and all other requirements are feasible, and zero if every $D_q\geq1/2$.
\end{proposition}

Restricting the permitted tuples to $(0,\ldots,0)$ and $(1,\ldots,1)$ leaves both minimum rates unchanged. These are exactly the joint predictions obtainable from the readouts $f_q(z,k,n)=z$ and $h_q(y)=y$. The reduction from the full product $\{0,1\}^m$ to these two tuples therefore lets us identify the optimized prediction rates with the source and split-feature rates for these readouts. Direct-feature coding has the same rate as source coding, $R_{XY}(\mathbf D)=R_X(\mathbf D)$, because it can reconstruct either of the two feature values.

Both task content and tolerated risk determine which profiles enter the equality condition. At a constant optimum, tasks with normalized allowance above $e^*$ have slack risk constraints and zero multipliers. If every task permits the same fraction $\alpha\in(0,1)$ of the interval from its Bayes risk to the constant-prediction risk, then $D_q=p_q+\alpha(1/2-p_q)$, $e^*=\alpha/2$, and every task is active. Rate equality then requires $\mathbf 1$ to belong to the convex hull of the entire family. Repeating a task changes neither its constraint nor this convex hull.

\paragraph{The full joint prediction alphabet.}
For Proposition~\ref{prop:confidence}, write $r_q=D_q-p_q$ for the permitted excess risk. For a joint prediction $\hat t\in\{0,1\}^m$, task $q$ has conditional excess distortion $w_q(k)\mathbf1\{\hat t_q\ne z\}$. We can remove independent $N$ by averaging the coding channel conditionally on $(Z,K)$, without changing any task risk or increasing mutual information. For nonnegative multipliers, put $L_k=\sum_q\lambda_qw_q(k)$ and $u_k(\hat t)=\sum_q\lambda_qw_q(k)\hat t_q$. By the variational formula in \eqref{eq:observation-variational}, we minimize
\begin{equation}
-\mathbb E_{Z,K}\log_2\sum_{\hat t\in\{0,1\}^m}
\pi(\hat t)2^{-\sum_q\lambda_qw_q(K)\mathbf1\{\hat t_q\ne Z\}}
\label{eq:confidence-full-output-dual}
\end{equation}
over probability distributions $\pi$ on all binary prediction tuples. To obtain the constrained dual objective, subtract $\sum_q\lambda_qr_q$.

The objective is convex in $\pi$. Complementing $Z$ and all predictions preserves the source law and costs, so averaging $\pi$ with its complement cannot increase the objective. A complement pair of total mass $\theta$ then contributes
\begin{equation}
\frac{\theta}{2}\left[2^{-u_k(\hat t)}+2^{-(L_k-u_k(\hat t))}\right]
\le\frac{\theta}{2}(1+2^{-L_k})
\label{eq:confidence-pair-partition}
\end{equation}
to the partition sum for either source bit. For $0\le u\le L_k$, the inequality follows from
\begin{equation}
1+2^{-L_k}-2^{-u}-2^{-(L_k-u)}
=(1-2^{-u})(1-2^{-(L_k-u)})\ge0.
\label{eq:confidence-pair-inequality}
\end{equation}
We can therefore assign each pair's probability to the all-zero and all-one pair without decreasing any partition sum. The argument applies separately at every confidence state. This extends the single-task symmetrization of \citet[Appendix I]{martinian2008distortion} without requiring symmetry among tasks or confidence states. Multiple-distortion Gibbs channels also appear in \citet{stavrou2023fidelity}.

Both the full-product and common-prediction Lagrangian infima are consequently
\begin{equation}
1-\mathbb E\log_2(1+2^{-L_K}),
\label{eq:confidence-diagonal-dual}
\end{equation}
attained by crossover probabilities $e(k)=1/(1+2^{L_k})$. When every $r_q>0$, mixing the exact common prediction with a sufficiently small positive probability of uniform joint predictions gives full support and strictly satisfies every risk constraint. In the problem restricted to a common prediction, a sufficiently small positive constant crossover probability lies in $(0,1/2)$ and strictly satisfies every risk constraint. Slater's condition holds for both convex programs. Their equal Lagrangian infima for every nonnegative multiplier vector imply equal constrained optima.

If some $r_q=0$, positivity of $w_q(k)$ forces $\widehat T_q=Z$, and data processing requires at least one bit. Sending $Z$ uses one bit per sample and gives every task zero excess risk, satisfying all nonnegative excess-risk allowances. Negative allowances are infeasible. If every $D_q\ge1/2$, a constant common prediction gives zero rate.

\paragraph{Error allocation and restricted observations.}
Let $U$ be the common binary prediction, so $\widehat T_q=U$ for every task. Averaging the channel with its bit-complemented counterpart gives a binary symmetric channel conditional on $K$. Its output is fair and independent of $K$, so
\begin{equation}
I(Z,K;U)=1-\mathbb Eh_2(e(K)),\qquad
\mathbb E d_{T_q}(T_q,U)=p_q+\mathbb E[w_q(K)e(K)].
\label{eq:confidence-common-channel}
\end{equation}
Replacing any conditional crossover probability above one half by its complement preserves its entropy and reduces every task risk. This proves Eq.~\ref{eq:confidence-source-rate} on the stated interval. When the encoder observes only $Y=Z$, the conditional error weights are $\bar w_q$. The same complement-pair argument proves optimality of a common prediction. Its tightest normalized requirement gives Eq.~\ref{eq:confidence-feature-rate}.

\paragraph{The active-profile criterion.}
Suppose $0<e^*<1/2$. If nonnegative coefficients $\rho_q$, indexed by $q\in J$, sum to one and satisfy $\sum_{q\in J}\rho_q\mathbf v_q=\mathbf 1$, the same weighted sum of the normalized risk constraints gives $\mathbb Ee(K)\leq e^*$. Jensen's inequality and monotonicity of binary entropy on $[0,1/2]$ give
\begin{equation}
\mathbb Eh_2(e(K))\leq h_2(\mathbb Ee(K))\leq h_2(e^*).
\label{eq:confidence-constant-optimal}
\end{equation}
The constant allocation $e(k)=e^*$ is feasible, and both inequalities hold with equality.

Conversely, the entropy objective is strictly concave, and $e^*$ is inside the error interval. The constant allocation $e^*/2$ strictly satisfies every constraint. The Karush--Kuhn--Tucker conditions are therefore necessary and sufficient. At the constant optimum, the constraints outside $J$ have strict slack and zero multipliers. Stationarity gives
\begin{equation}
h_2'(e^*)\mathbf 1=\sum_{q\in J}\mu_q\mathbf v_q,\qquad \mu_q\geq0.
\label{eq:confidence-kkt}
\end{equation}
All confidence-state probabilities are positive, so they cancel from the coordinate equations. Taking expectation over $K$ gives $\sum_{q\in J}\mu_q=h_2'(e^*)>0$. Dividing each multiplier by this sum gives the required convex combination. If $\mathbf 1$ is outside that convex hull, the constant allocation fails a necessary optimality condition. The feasible set is compact, so a maximizing allocation exists and has $\mathbb E h_2(e(K))>h_2(e^*)$. Hence the rate from $X$ is strictly smaller.

\paragraph{Two-state and three-state families.}
For two equiprobable confidence states and first-task weights $(w_0,w_1)$, Eq.~\ref{eq:confidence-source-rate} reduces to the single-task expression of \citet[Section IV-E, Eqs.~40--44]{martinian2008distortion},
\begin{equation}
R_X^{(1)}(D)=1-\max_{e_0,e_1}\frac{h_2(e_0)+h_2(e_1)}{2},
\label{eq:confidence-two-state}
\end{equation}
where $e_k=e(k)$ and $p$ is the task's Bayes risk. We maximize over $e_0,e_1\in[0,1/2]$ satisfying
\[
\frac{w_0e_0+w_1e_1}{2}\leq D-p.
\]
The Gibbs crossover probabilities satisfy $e_k=1/(1+2^{\lambda w_k})$. Swapping the weights for the second task makes the two normalized profiles average to $\mathbf 1$. At equal tolerances, the active-profile criterion gives Eq.~\ref{eq:confidence-complementary-rate}. For a single nonconstant profile, its convex hull does not contain $\mathbf 1$, so $R_X^{(1)}(D)<R_Y^{(1)}(D)$ for $p<D<1/2$.

For three equiprobable confidence states, choose three tasks with rows of conditional error costs
\begin{equation}
\begin{pmatrix}
0.9&0.5&0.1\\
0.1&0.9&0.5\\
0.5&0.1&0.9
\end{pmatrix}.
\label{eq:confidence-three-state}
\end{equation}
Since $\epsilon_q(k)=(1-w_q(k))/2$, each task's noise probabilities are a permutation of $(0.05,0.25,0.45)$. Their mean is the Bayes risk $p_q=0.25$. The chosen component tolerance $D_q=0.30$ leaves excess risk $0.05$. Using Eq.~\ref{eq:confidence-source-rate}, we maximize $\frac13\sum_{k=1}^3h_2(e_k)$ subject to $\sum_{k=1}^3w_q(k)e_k\le0.15$ for every included task. The numerical optima are
\begin{center}
\appendixtableformat
\begin{tabular*}{\linewidth}{@{\hspace{3pt}\extracolsep{\fill}}lcc@{\hspace{3pt}}}
\toprule
Required tasks & Optimal $(e_1,e_2,e_3)$ & $R_X$ (bits per sample)\\
\midrule
\appendixtableband{1}
First & $(0.0398,0.1458,0.4125)$ & $0.394$\\
First two & $(0.1210,0.0423,0.1996)$ & $0.498$\\
\appendixtableband{1}
All three & $(0.1,0.1,0.1)$ & $0.531$\\
\bottomrule
\end{tabular*}
\end{center}
Cyclic permutation gives the same rates for every singleton or pair. With all three rows, summing the constraints bounds mean error by $0.1$, attained by the constant allocation. The three normalized profiles average to $\mathbf1$, which lies on no pair's segment. For an encoder observing only $Y=Z$, the error probability $e$ is independent of $K$ and each task's mean error weight is $0.5$. Thus $0.5e\le0.05$ gives $R_Y=1-h_2(0.1)\approx0.531$ for every nonempty task subset. Every task's uncompressed Bayes prediction remains $Z$.

\section{Empirical methods and supporting evidence}
\label{app:empirical}
\subsection{Finite-model numerical comparisons}
\label{app:exact-computations}

We separately minimize mutual information over the full product prediction alphabet and over common-prediction error allocations. Sixteen cases include cyclic subsets, unequal normalized allowances, nonuniform confidence probabilities and families with up to four tasks. The two minima differ by at most $4.84\times10^{-8}$ bits. We also test the whole-family and active-family convex-hull conditions separately.

\subsection{Three-class operational coding}
\label{app:nonbinary-codes}

The law in \eqref{eq:nonbinary-reference-law} has five distinct conditional-cost rows because the confidence states at $Z=2$ have identical costs. Each codebook starts from 8,192 distinct ternary words of length 16. Three Lloyd updates assign 131,072 training blocks by total conditional loss and replace each word position with its minimum-loss class. We use 16,384 development blocks for the first fit and 32,768 for each replication, and fix the additional-task tolerance at $0.499$ before independent confirmation.

Known-loss encoding minimizes the sum of additional conditional losses. Fitted-loss encoding estimates five rows from $2^{20}$ independent observations with noisy targets sampled per codebook. For reduced-observation encoding, we pool these estimates by class. We round fitted costs to multiples of $2^{-16}$ for integer selection scores, with per-entry error at most $2^{-17}$. Each confirmation packet contains $2^{21}$ 13-bit indices, padding, and a 48-byte header, totaling 3,407,920 bytes for 33,554,432 source symbols. The shared codebook and packet suffice for decoding.

We assess the codes by averaging conditional target losses within independent fresh blocks. For $N$ blocks and loss range $b$, the one-sided Hoeffding allowance is $b\sqrt{\ln(22/0.01)/(2N)}$. A union bound gives simultaneous 99\% coverage for 22 component means across the known and fitted full selectors, fitted reduced selectors, and two controls. The original loss range is one, and the additional range follows from the source table. The largest full-observation upper risk bounds are $0.25248$ (original task) and $0.49846$ (additional task).

The original-task control minimizes class errors. The confidence-removed control minimizes additional loss averaged over $K$, retaining the marginal target law and unique Bayes prediction. Each uses one fixed codebook and 32,768 paired blocks. Full and reduced encoders produce identical packets. The confidence-removed control has additional risk $0.51689$, above the $0.499$ tolerance.

\subsection{Binary learned block coding}
\label{app:block-coding}

\subsubsection{Source laws, fitting, and messages}

Decision bits and confidence states are independent and fair. The first task has noise probabilities $(0.05,0.45)$ and tolerated risk $0.30$. The additional task is a duplicate or has reversed noise probabilities, with tolerance $0.46$ (loose) or $0.30$ (tight). Confidence is observed through independent bit corruption at probability $\tau\in\{0,1/8,1/2\}$. The first-task conditional error costs become $(0.9-0.8\tau,0.1+0.8\tau)$. Three-state profiles are the rows of Eq.~\ref{eq:confidence-three-state}.

We use 4--10 index bits for length 12 and 5--13 for length 16. Eight weighted Lloyd rounds start from distinct random words, assigning 65,536 fresh blocks per round and updating bits by weighted majority. Fitting costs are uniform, original-task costs, or the first-two-profile average for three states. We compare both observations for each frozen codebook and select across fitting origins. All decoders return the same word.

Reference encoding minimizes known conditional loss. Fitted encoding learns two or three positive state scores, normalized to the same total cost, from soft known-cost assignments on a separate 64-word, length-12 codebook, using 250 batches of 512 independent blocks. Selection is exhaustive. Reduced-observation encoding uses constant scores and independent dummy states.

Other second-task limits are $0.34$, $0.38$, and $0.42$. Corrupted-state comparisons retain the single and tight-pair requirements. We set each task's tolerance to $0.30$ in the three-state single, first-two, and cyclic families.

We generate candidate assignments from scalarized profiles and minimize mean validation payload rate over their nonnegative frequencies, with each component risk $0.0015$ below its tolerance. We convert frequencies to integer block counts in a shuffled, source-independent schedule shared with the decoder. For candidate $c$, we transmit an eight-byte header and packed $b_c$-bit indices for $N_c>0$ blocks. The complete rate is
\begin{equation}
r=\frac{8}{n\sum_c N_c}\sum_c
\left(8+\left\lceil\frac{N_cb_c}{8}\right\rceil\right)
\quad\text{bits per decision}.
\label{eq:block-complete-rate}
\end{equation}
Codebooks and the source-independent schedule are shared with the decoder.

\subsubsection{Assessment and score precision}
\label{app:score-precision}

Each of three fits uses 65,536 validation and 131,072 assessment blocks, separate from fitting and calibration. We use independent-block standard errors and normal multipliers $1.96$ for one requirement and $2.394$ for two or three to form intervals with approximate simultaneous 95\% coverage conditional on fitting and selection. The rate-equivalence margin is $1/64$ bit per decision.

Fitted-score differences of order $10^{-7}$ alter assignments at exact known-cost ties. We round fitted scores to five decimal places, retain parameters and codebooks, and use fresh validation, assessment, and forecast samples.

\subsubsection{Finite-code sensitivity and fixed-codebook comparisons}

\begin{figure}[t]
\centering
\includegraphics[width=\linewidth]{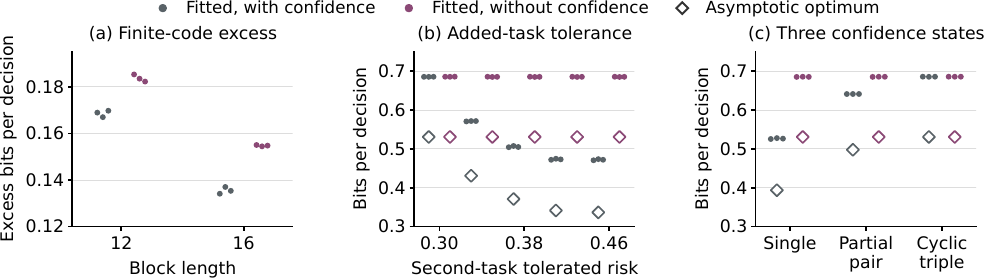}
\caption{Finite-code sensitivity. Filled circles show three fits per condition and hollow diamonds the unrestricted asymptotic rates. (a) Single-task rate above its asymptotic minimum at lengths 12 and 16. (b) Length-16 rates as the second-task limit varies, with the first fixed at $0.30$. (c) Three-state families at length 16. Horizontal offsets separate fits. We compute rates from index, header and padding bits.}
\label{fig:block-sensitivity}
\end{figure}

The single-task saving persists at both block lengths (Figure~\ref{fig:block-sensitivity}). Single-task, duplicate, and loose-pair rates overlap. Corrupting confidence reduces the saving until it vanishes at $\tau=1/2$, while tight-pair rates remain equivalent at all three corruption levels. With exact confidence, single-task coding concentrates errors in the state with lower conditional error cost. When both complementary requirements are tight, the two conditional error probabilities are nearly equal.

As three-state profiles are added, the full-observation rate increases. For the cyclic family, the full- and reduced-observation rates are equivalent, while the optimal uncompressed predictions remain unchanged. Across both lengths and all conditions, fitted and reference rates differ by at most $0.000448$ bits per decision. For all 252 selected policies, the joint upper risk bounds are at or below their respective task tolerances.

On fixed length-16 codebooks with eight-bit indices, observing confidence reduces single-task risk by $0.03779$--$0.03802$, with block standard error $0.000105$. These codebooks coincide across fitting origins. With ten-bit indices, full-observation risks span $0.272730$--$0.272888$ across both fitting origins, compared with $0.308054$--$0.308321$ without confidence. The advantage therefore persists on codebooks fitted for either observation. For the constant-score complementary candidate, predictions and component risks agree exactly across observations on each codebook.

\subsubsection{Additional-task risk forecasts}

For a fixed binary code with conditional error probabilities $e(k)$, another known task has risk
\begin{equation}
L_q=p_q+\sum_k\Pr(K=k)w_q(k)e(k).
\label{eq:measured-profile-risk}
\end{equation}
We substitute conditional error probabilities estimated on 131,072 calibration blocks into \eqref{eq:measured-profile-risk} to forecast additional-task risks for each frozen single-task code without refitting. Across 252 forecasts, the largest absolute component-risk error on independent assessment is $0.001128$. Forecast and assessment classifications agree for both the mean-threshold and joint-interval rules, including the two uncertain cases at tolerance $0.42$. The length-16 codes have complementary-task risk of about $0.420$, below the loose tolerance and above the tight tolerance, with unchanged messages. Stricter requirements can therefore require another code even though the uncompressed Bayes predictions agree.

\subsection{ImageNet task requirements}
\label{app:imagenet-methods}

ENTITY-30 contains 307,828 training images across 240 fine classes \citep{santurkar2021breeds}. We split official validation data into 6,000 validation and 6,000 test images, each with 25 per fine class. Preprocessing uses a 232-pixel resize, 224-pixel center crop and channel normalization. We normalize the selected 240 logits into fine probabilities, then sum them within each superclass for coarse probabilities.

The $512\times28\times28$ input tensor is normalized using 4,096 training images. Analysis widths are $512,192,64$. We pad the spatial dimensions to $32\times32$, then compute a $64\times8\times8$ latent and $32\times2\times2$ hyperlatent with the analysis transforms. Synthesis restores the input dimensions. We compute complete rates from the bits used for both entropy-coded latents and 40 framing bytes per image.

We fit the common initializer by minimizing normalized feature MSE plus $0.003$ times estimated bpp with Adam at $10^{-3}$. Task fits use fresh Adam optimizers at $10^{-4}$. Each stage uses 5,000 updates, batches of 128 and 639,976 image exposures, with matched image order across objectives. Across both objectives, we retain all eight final checkpoints from fits using seeds 17 and 23 and rate multipliers $0.03$ and $0.3$. Coarse KL and fine cross entropy are normalized by $\ln30$ and $\ln240$. Neither objective includes feature MSE.

Adapted readouts use 2,048-dimensional penultimate suffix features, a 512-unit hidden layer and separate coarse and fine outputs. We fit two readouts with different seeds on the same 23,040 training images, 96 per fine class, for 20 AdamW epochs at learning rate $10^{-3}$, weight decay $10^{-4}$ and batch size 256. On validation, we select checkpoints by summed normalized cross entropy and each task's route by comparing inherited accuracy with the two-fit adapted mean, for raw and coded inputs alike. We average rates and accuracies over the two codec fits per objective and multiplier. We test the union of validation selections from Section~\ref{sec:imagenet}.

We calculate intervals from 10,000 common paired resamples within fine classes, conditional on fitted models and classes. One-sided 97.5\% upper limits give approximate 95\% Bonferroni coverage for each group's coarse- and fine-accuracy drops. Validation sensitivity comparisons use fine allowances of three and ten points.

The selected added-fine code's simultaneous upper drop limits are $2.575$ coarse and $4.367$ fine points, below the three- and five-point allowances.

\subsection{CIFAR-100}
\label{app:cifar-methods}

\subsubsection{Fitting and complete rates}

We reserve 250 training images per coarse class for validation, leaving 45,000 for fitting. The 10,000 test images are excluded from fitting. Earlier test results had been inspected. For the extension in Section~\ref{sec:cifar}, we select candidates on validation. We exclude fine labels from data splitting and teacher selection.

The three ResNet-18 teachers use a $3\times3$, stride-one input convolution, no max pooling, independent initializations and 200 epochs with padded random crops and reflections. We select checkpoints by validation coarse accuracy, breaking ties by cross entropy.

Early-observation and logit-observation codecs use strided convolutional and MLP analysis, respectively, with $64\times4\times4$ latents, $32\times1\times1$ hyperlatents and common transposed-convolution synthesis. Reconstructions are unconstrained. A mean-and-scale hyperprior models conditional Gaussian main latents and factorized hyperlatents without autoregressive context \citep{balle2018variational,minnen2018joint,begaint2020compressai}.

Preservation MSEs are normalized by training constant-predictor errors. For codec fitting, we use fine labels only in the family-supervised control, averaging coarse and fine cross entropies normalized by training-label entropies. Fine predictions are computed by a frozen early-feature MLP. We train randomly initialized codecs for 60 epochs. We fit paired preservation codecs for 60 epochs from a shared codec pretrained for 60 epochs with early-feature MSE. The earlier three-fit Teacher A groups and six reused controlled fits use final epoch-60 checkpoints. For single-fit controls, the reused warm control and new extension fits, we select checkpoints minimizing validation distortion plus $\beta$ times estimated bpp.

Fine readouts are linear classifiers or MLPs with two 512-unit hidden layers on tensors pooled to $4\times4$ or on 20 logits. We normalize coordinates using training data, fit for 100 epochs and select checkpoints by validation accuracy. Training inputs are deterministically quantized, and assessment inputs are entropy-decoded. We compute rates from both entropy-coded bitstream lengths and the 40-byte header ($0.3125$ bpp).

\subsubsection{Preservation selection and test assessment}

Each paired candidate uses two codecs per teacher and two MLP readouts per codec and input type, with matched examples and fitting rules. The earlier reconstructed-logit diagnostic uses a fixed raw-logit classifier, refitted in the extension.

We require an additional matched readout or codec pair when validation ranges across readout fits or codec means exceed one point and could affect selection. For candidates that could affect selection, readout ranges are at most $0.88$ point.

We compare 14 paired groups and six single-fit controls (34 codecs) on validation. For each teacher and loss, we select the paired candidate minimizing mean rate subject to both mean accuracy-drop allowances, using two refitted tensor readouts.

\begin{figure}[t]
\centering
\includegraphics[width=\linewidth]{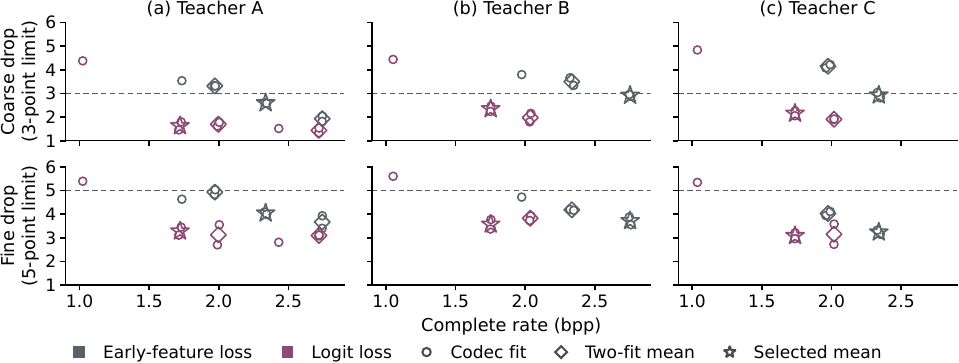}
\caption{CIFAR validation candidates by teacher. Dashed lines mark the coarse and fine allowances. Open circles show codec fits, diamonds paired means, and stars the six selections. Single-fit controls have no paired mean. Fine accuracies average two refitted tensor readouts.}
\label{fig:cifar-validation}
\end{figure}

\begin{table}[t]
\centering
\appendixtableformat
\caption{Uncompressed CIFAR-100 reference accuracies (\%). Fine-readout means and ranges summarize two fits. We use the early-tensor mean as the fine-accuracy reference.}
\label{tab:cifar-extension-references}
\begin{tabular*}{\linewidth}{@{\hspace{3pt}\extracolsep{\fill}}lcrrrrr@{\hspace{3pt}}}
\toprule
& & & \multicolumn{2}{c}{Fine, early tensor} & \multicolumn{2}{c}{Fine, logits}\\
\cmidrule(lr){4-5}\cmidrule(l){6-7}
Split & Teacher & Coarse & Mean & Range & Mean & Range\\
\midrule
\appendixtableband{3}
Validation & A & 85.68 & 63.370 & [63.36, 63.38] & 45.970 & [45.84, 46.10]\\
& B & 85.86 & 62.810 & [62.40, 63.22] & 46.410 & [46.38, 46.44]\\
& C & 86.16 & 62.430 & [62.30, 62.56] & 45.390 & [45.38, 45.40]\\
\addlinespace[3pt]
Test & A & 85.38 & 62.615 & [62.43, 62.80] & 44.485 & [44.28, 44.69]\\
& B & 85.27 & 62.480 & [62.46, 62.50] & 45.060 & [44.65, 45.47]\\
& C & 85.55 & 63.060 & [63.05, 63.07] & 43.765 & [43.67, 43.86]\\
\bottomrule
\end{tabular*}
\end{table}

We test the six selected groups and three feature-preserving candidates at the next lower rates, fixed using validation data (18 codecs, Fig.~\ref{fig:cifar-local}). References are in Table~\ref{tab:cifar-extension-references}. We calculate paired 95\% intervals from 2,000 common bootstrap resamples within fine classes, averaging per-image correctness across fits, conditional on teachers, codecs and readouts. For an approximate 95\% Bonferroni joint assessment of each group's mean risks, both upper endpoints must be at or below their respective allowances.

The selected feature-preserving group for Teacher C and the feature-preserving candidates with the next lower rates for teachers A and C fail the joint assessment, with coarse-drop intervals $[2.310,3.135]$, $[2.470,3.365]$ and $[2.805,3.705]$ points. Teacher C's selected group and the candidate with the next lower rate contain one and two failed codec fits, respectively. Other assessed mean accuracy drops are within the allowances.

Intervals for the selected logit-minus-feature fine-accuracy differences for teachers A, B and C are $[-0.640,0.280]$, $[-0.835,0.055]$ and $[0.050,0.928]$ points, within the one-point equivalence margin. Corresponding rate reductions are $26.35\%$, $36.38\%$ and $25.73\%$.

\subsubsection{Observation and initialization controls}
\label{app:cifar-development}
\label{app:cifar-test}

The earlier Teacher A comparison uses three codec fits and one fine-readout initialization. Randomly initialized early-observation and logit-observation codecs have test fine accuracies of $61.43\%$ and $41.22\%$ at $2.770$ and $0.724$ bpp with refitted readouts. Their different analysis architectures prevent isolating the observation cost. We fitted the early-observation logit-loss codec without early-feature MSE pretraining.

Earlier paired preservation fits have $59.88\%$ fine accuracy at $2.750$ bpp with early-feature loss and $60.14\%$ at $1.999$ bpp with logit loss. Logit-loss fits have $3.4812$ times the early-feature MSE of the early-feature-loss fits. At a separate near-rate logit-loss setting, refitting improves fine accuracy from $40.89\%$ to $60.27\%$.

Another randomly initialized logit-loss fit has validation coarse accuracy of $66.00\%$, compared with the $85.68\%$ reference. Fine-supervised controls have lower coarse accuracy than the focal settings, so we cannot infer an advantage at matched coarse risk.

\subsection{Taskonomy}
\label{app:taskonomy-methods}

\subsubsection{Data, losses, and fitting}

We use $256\times256$ images, with 24,000 training images balanced across 24 buildings, 500 validation images from five other buildings, and 2,000 test images balanced across five further buildings. All candidates use convolutional hyperprior codecs and width-48 residual upsampling decoders.

Depth loss is unit-transition Smooth-L1 between log predictions and log targets, with depths clamped to at least $10^{-4}$ before taking logarithms. Semantic targets are Taskonomy's FCIS pseudo-labels.\footnote{\url{https://github.com/StanfordVL/taskonomy/blob/master/data/README.md}} We use 17 labels including background and exclude uncertain pixels. Cross-entropy weights are inverse square roots of class frequencies in 2,400 balanced training images. We divide each image's depth loss by its valid-pixel count and semantic loss by its valid-pixel weight sum, then average across images. During training, we normalize over valid pixels within each batch. We compute object mIoU over non-background classes present in the ground truth. Edge targets are clipped, rescaled Sobel magnitudes of Gaussian-smoothed grayscale images derived from RGB. Edge and RGB losses are MSE.

We fit each codec for 24,000 updates with batches of 12 images, minimizing estimated bpp times a multiplier plus task losses weighted by 10, 1, 10, and 100 for depth, semantics, edges, and RGB. After freezing codecs, we fit decoder sets with equal exposure and select checkpoints by weighted-$DSE$ validation loss. For available-decoder results, we select each task's original or retained newly fitted width-48 decoder on validation. For matched-decoder results, we use the common weighted-sum checkpoint. We fix decoders before testing and use one message per decoding procedure.

Absolute requirements are fixed from validation reference and constant-predictor losses,
\begin{equation}
D_q(\alpha)=L_{q,\mathrm{ref}}+
\alpha(L_{q,\mathrm{const}}-L_{q,\mathrm{ref}}),
\qquad \alpha\in\{0.1,0.25,0.5\}.
\label{eq:empirical-tolerance}
\end{equation}
We fit the uncompressed reference with the same architecture families and exposure, without quantization or rate penalty. The constant predictions are geometric-mean depth, weighted smoothed class probabilities, and mean edge intensity, computed from training data. For constant depth RMSE, we use arithmetic-mean depth. Additional mIoU floors are $(1-\alpha)$ times validation reference mIoU, giving $0.16573044$, $0.13810870$, and $0.09207246$, with cross-entropy requirements retained.

\subsubsection{Candidate selection and fit variation}
\label{app:taskonomy-choices}

We fit one codec and its task decoders for each of twelve settings. We select $D,30$, $DS,3$, and $DE,3$ for three-fit replication by mean validation rates and risks. We compute pointwise 95\% intervals from 2,000 paired image resamples within the five test buildings, conditional on buildings and fitted models. Matched newly fitted decoders give the same descriptive test minima as the available-decoder rule in Table~\ref{tab:taskonomy-pool-test}.

\begin{table}[t]
\centering
\appendixtableformat
\caption{Complete Taskonomy test pool with task decoders selected on validation. We compute rates from complete message lengths and exclude background from object mIoU. Depth RMSE is in metres.}
\label{tab:taskonomy-pool-test}
\begin{tabular*}{\linewidth}{@{\hspace{3pt}\extracolsep{\fill}}lrrrrrrr@{\hspace{3pt}}}
\toprule
Family & Multiplier & Rate & Depth & Semantics & Edges & Object & Depth\\
& & (bpp) & $L_D$ & $L_S$ & $L_E$ & mIoU & RMSE (m)\\
\midrule
\appendixtableband{3}
D & 0.3 & 0.07096 & 0.07669 & 0.84068 & 0.02827 & 0.0444 & 1.251\\
& 3 & 0.01752 & 0.07551 & 0.92659 & 0.03642 & 0.0284 & 1.223\\
& 30 & 0.00718 & 0.08833 & 1.03715 & 0.04183 & 0.0082 & 1.350\\
\addlinespace[3pt]
DS & 0.3 & 0.12416 & 0.07607 & 0.66453 & 0.02421 & 0.1303 & 1.228\\
& 3 & 0.03346 & 0.07688 & 0.66444 & 0.03263 & 0.1345 & 1.257\\
\addlinespace[3pt]
\appendixtableband{3}
DE & 0.3 & 0.13277 & 0.07419 & 0.78960 & 0.00299 & 0.0591 & 1.201\\
& 1 & 0.06386 & 0.07250 & 0.82183 & 0.00402 & 0.0485 & 1.195\\
& 3 & 0.03142 & 0.07703 & 0.87632 & 0.00758 & 0.0367 & 1.260\\
\addlinespace[3pt]
DSE & 0.3 & 0.19188 & 0.07777 & 0.62568 & 0.00425 & 0.1384 & 1.248\\
& 1 & 0.09410 & 0.07364 & 0.62297 & 0.00543 & 0.1441 & 1.208\\
& 3 & 0.05095 & 0.07676 & 0.65327 & 0.00770 & 0.1204 & 1.265\\
\addlinespace[3pt]
\appendixtableband{1}
RGB & 0.3 & 0.18272 & 0.07755 & 0.67093 & 0.00055 & 0.1262 & 1.211\\
\midrule
Reference & -- & -- & 0.07375 & 0.62329 & 0.00189 & 0.1361 & 1.184\\
\bottomrule
\end{tabular*}
\end{table}

With the mIoU floor, we select $DS,3$ and $DSE,3$ on validation for loose $DS$ and $DSE$ requirements. No candidate satisfies stringent test requirements including semantics. Both selected codes and the uncompressed reference have test object mIoU below the moderate floor.

\begin{table}[t]
\centering
\appendixtableformat
\caption{Taskonomy test means [ranges] over three fits with available-decoder selection. All fits satisfy the moderate depth requirement. The last row counts fits satisfying moderate $DS$, $DE$, and $DSE$ requirements.}
\label{tab:taskonomy-seeds-test}
\setlength{\tabcolsep}{3pt}
\begin{tabular*}{\linewidth}{@{\hspace{3pt}\extracolsep{\fill}}lrrr@{\hspace{3pt}}}
\toprule
Metric & $D,30$ & $DS,3$ & $DE,3$\\
\midrule
\appendixtableband{1}
Rate (bpp) & 0.00725 [0.00718, 0.00736] & 0.03199 [0.03077, 0.03346] & 0.02808 [0.02429, 0.03142]\\
Depth $L_D$ & 0.08963 [0.08833, 0.09099] & 0.07598 [0.07543, 0.07688] & 0.07706 [0.07675, 0.07740]\\
\appendixtableband{1}
Semantics $L_S$ & 1.05059 [1.03715, 1.05881] & 0.64969 [0.63882, 0.66444] & 0.88166 [0.87632, 0.88692]\\
Edges $L_E$ & 0.04171 [0.04155, 0.04183] & 0.03232 [0.03180, 0.03263] & 0.01054 [0.00758, 0.01439]\\
\appendixtableband{1}
Object mIoU & 0.00864 [0.00819, 0.00917] & 0.13315 [0.12376, 0.14116] & 0.03844 [0.03589, 0.04273]\\
Depth RMSE (m) & 1.355 [1.350, 1.364] & 1.255 [1.251, 1.257] & 1.271 [1.260, 1.282]\\
\midrule
Fits: $DS/DE/DSE$ & 0/0/0 & 3/0/0 & 2/2/1\\
\bottomrule
\end{tabular*}
\end{table}

In the hardest building, all fits exceed the moderate depth-loss tolerance. For every building and fit, $DS$ has lower semantic and higher edge risk than $DE$. We double newly fitted decoder width from 48 to 96 for the three base fits, keeping fitting exposure and selection unchanged. Primary validation and test feasibility decisions are unchanged.

\subsubsection{Input-independent mixtures}
\label{app:taskonomy-mixtures}

We select frequencies $a_c$ for a shared image-independent schedule by minimizing validation rate subject to the required task risks
\begin{equation}
\min_{a}\sum_c a_c r_{c,\mathrm{val}},
\qquad
a_c\ge0,\quad\sum_c a_c=1,\qquad
\sum_c a_c L_{cq,\mathrm{val}}\le D_q
\quad\text{for each required }q.
\label{eq:taskonomy-mixture}
\end{equation}
We use the twelve original candidates and validation-selected readouts, then apply weights unchanged on test. The encoder and decoder share the schedule, so no per-image codec identifier need be transmitted.

\begin{table}[t]
\centering
\appendixtableformat
\caption{Validation-selected codes and image-independent mixtures. We count all transmitted bits. We select $D,30$ for depth only at every tolerance. Its test risk exceeds only the stringent tolerance.}
\label{tab:taskonomy-validation-choices}
\label{tab:taskonomy-mixtures}
\begin{tabular*}{\linewidth}{@{\hspace{3pt}\extracolsep{\fill}}cclrrrrl@{\hspace{3pt}}}
\toprule
& Required & Single & \multicolumn{3}{c}{Validation} & \multicolumn{2}{c}{Mixture test}\\
\cmidrule(lr){4-6}\cmidrule(l){7-8}
$\alpha$ & tasks & code & Single rate & Mixture rate & Reduction & Rate & Unsatisfied requirement\\
\midrule
\appendixtableband{3}
0.10 & $DS$ & $DS,3$ & 0.03376 & 0.03123 & 7.50\% & 0.03095 & None\\
& $DE$ & $DE,1$ & 0.06241 & 0.03791 & 39.26\% & 0.03837 & Edge\\
& $DSE$ & $DSE,1$ & 0.09171 & 0.04897 & 46.60\% & 0.04991 & Edge\\
\addlinespace[3pt]
0.25 & $DS$ & $DS,3$ & 0.03376 & 0.02459 & 27.18\% & 0.02436 & None\\
& $DE$ & $DE,3$ & 0.03123 & 0.02692 & 13.81\% & 0.02706 & Edge\\
& $DSE$ & $DSE,3$ & 0.04994 & 0.03199 & 35.95\% & 0.03206 & Edge\\
\addlinespace[3pt]
\appendixtableband{2}
0.50 & $DS$ & $D,3$ & 0.01745 & 0.01351 & 22.56\% & 0.01338 & None\\
& $DE,DSE$ & $DE,3$ & 0.03123 & 0.01889 & 39.50\% & 0.01895 & Edge\\
\bottomrule
\end{tabular*}
\end{table}

Five of twelve mixtures and eleven of twelve single-code selections satisfy their test requirements. For all six mixtures with an edge requirement, the test edge risks exceed the tolerance. The three $DS$ mixtures and the depth-only choices at moderate and loose tolerances satisfy their test requirements.

For the moderate $DS$ mixture, we use $D,30$ with frequency approximately $0.346$ and $DS,3$ with frequency approximately $0.654$. Test depth and semantic risks are $(0.080843,0.793438)$.

\begin{figure}[t]
\centering
\includegraphics[width=\linewidth]{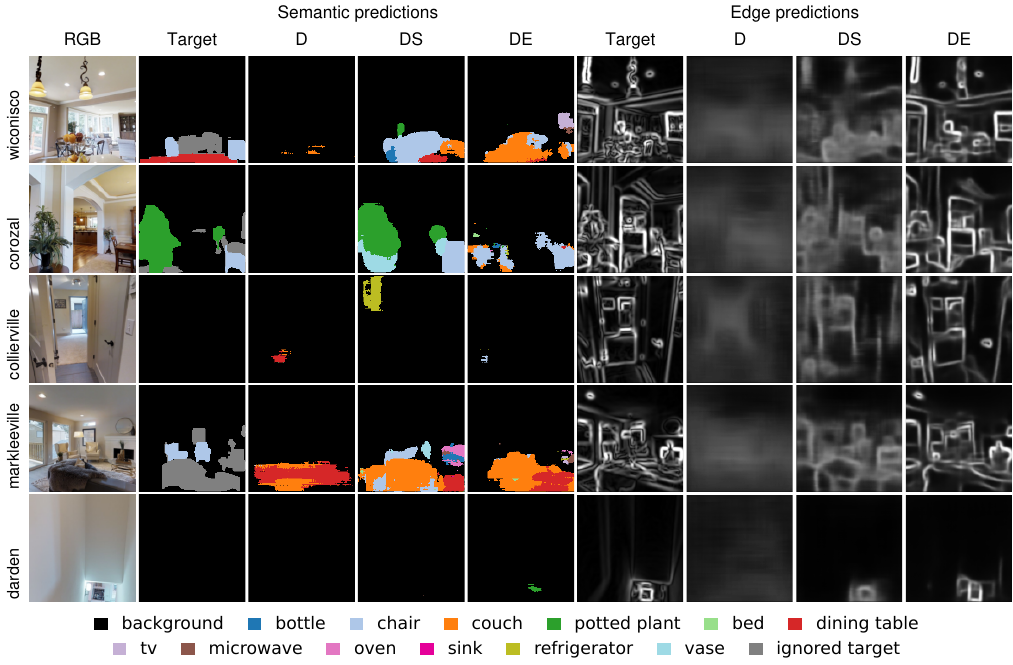}
\caption{Semantic and edge predictions from five decoded validation images. RGB references and class colors are identical across methods. Gray target pixels are excluded from loss. Edge intensity scale is $[0,1]$.}
\label{fig:taskonomy-examples}
\end{figure}

\subsubsection{Semantic class support}
\label{app:semantic-support}

Validation contains fourteen object classes. Test additionally contains \emph{book}, while \emph{toaster} is absent from both. Background occupies $92.7103\%$ and $93.5162\%$ of valid validation and test pixels. We compute original mIoU over object classes present in each split.

Restricting test mIoU retrospectively to the fourteen validation-present classes raises reference mIoU from $0.136058$ to $0.143284$ and mean $DS,3$ mIoU from $0.133147$ to $0.141463$, both above the unchanged moderate floor $0.13810870$. Two of three $DS$ fits have mIoU at or above this floor. The $DSE,3$ base fit has mIoU $0.127146$, below the floor. Test reference mIoU remains below the validation reference value of $0.184145$.

\subsubsection{Complete message lengths and decoded examples}
\label{app:taskonomy-bytes}
\label{app:qualitative}

Complete messages contain entropy-coded latents and a 40-byte header ($67.32\%$ of the depth-only message length). We compute rates from these lengths, excluding model storage and decoder computation.

We apply base $D,30$, $DS,3$, and $DE,3$ codes to the first validation image per building in Fig.~\ref{fig:taskonomy-examples}.

\subsection{Feature preservation and visual decoding}
\label{app:pilot}

We use ImageNet ResNet-50 \texttt{IMAGENET1K\_V2} features. Block 13 ends Stage 3 with a $1024\times14\times14$ tensor. Blocks 14--16 form Stage 4, each with a $2048\times7\times7$ tensor. The unchanged suffix reproduces the original ImageNet logits. At each depth, we input the same complete tensor to the inverse and native-feature classifier.

\paragraph{Photographs and species prediction.}
We square-crop CUB photographs around the annotated bird box, with side length $\lceil1.2\max(w,h)\rceil$, mid-gray boundary padding and bilinear resizing to $224\times224$ pixels \citep{wah2011cub}. We fit readers on 4,794 photographs from 200 species, validating on six per species. We evaluate species-pair accuracy on 30 official test photographs per species by restricting the fitted 200-class output to Indigo Bunting and Blue Grosbeak. Both displayed photographs belong to development.

Native readers use the pretrained fourth stage at block 13 and final residual block at blocks 14--16, followed by global pooling and a new 200-class linear output. All reader parameters are fitted with feature extraction fixed. For RGB readers, we adapt a pretrained ResNet-50 to each inverse's reconstructions, using horizontal reflections. We fit two readers per observation for 50 AdamW epochs with class-weighted cross entropy, learning rate and weight decay $10^{-4}$, and batch size 64. We select checkpoints by validation macro accuracy.

\paragraph{Visual inversion.}
Convolutional inverses use $5\times5$ kernels, 128 hidden channels, ReLUs, linear RGB outputs, and four upsampling layers at block 13 or five at blocks 14--16. We minimize RGB MSE for 100,000 Adam updates at learning rate $10^{-4}$, clipping gradient norms at one. We sample 64-image ImageNet training batches with replacement, shared across depths, and select checkpoints by MSE on 5,000 fixed ImageNet validation photographs. Pixels are clipped and rounded for eight-bit export. Using CUB part annotations, we locate enlarged $96\times48$ wing regions at identical coordinates across reconstructions.

\begin{table}[ht]
\centering
\appendixtableformat
\caption{Species-pair accuracy (\%) for each of the two fitted readers.}
\label{tab:pilot-pair}
\begin{tabular*}{\linewidth}{@{\hspace{3pt}\extracolsep{\fill}}lrrrr@{\hspace{3pt}}}
\toprule
Observation & Block 13 & Block 14 & Block 15 & Block 16\\
\midrule
\appendixtableband{2}
Native feature, fit 1 & 96.7 & 95.0 & 93.3 & 93.3\\
Native feature, fit 2 & 96.7 & 96.7 & 93.3 & 93.3\\
\addlinespace[3pt]
Reconstructed RGB, fit 1 & 91.7 & 86.7 & 88.3 & 83.3\\
Reconstructed RGB, fit 2 & 88.3 & 85.0 & 86.7 & 81.7\\
\bottomrule
\end{tabular*}
\end{table}

\end{document}